\pdfoutput=1

\documentclass[11pt, letterpaper]{article}
\usepackage[margin={2cm,2cm}]{geometry} 

\usepackage{sectsty}
\usepackage{abstract}
\usepackage{color}
\usepackage{graphicx}
\usepackage{multicol}
\usepackage{amsmath,amssymb,amstext,amsfonts}
\usepackage[usenames,svgnames]{xcolor}
\usepackage{subfig}
\usepackage{caption}
\usepackage{framed}
\usepackage{authblk}
\usepackage{titlesec}
\usepackage{setspace}
\usepackage{times,fancyhdr}
\usepackage[normalem]{ulem} 
\usepackage{enumerate} 
\usepackage{mathrsfs} 
\usepackage{bigints}
\usepackage{bm}
\usepackage{arydshln}
\usepackage{upgreek} 
\usepackage{enumitem} 
\usepackage[hypertexnames=false]{hyperref}
\usepackage{float}

\titlespacing*{\section}{0pt}{3ex plus 2ex}{1ex} 
\sectionfont{\fontsize{11}{11}\selectfont} 
\subsectionfont{\fontsize{10}{10}\selectfont} 

\hypersetup{
    colorlinks=true,
    urlcolor=SteelBlue,
    linkcolor=red,
    citecolor=blue,
}

\DeclareMathAlphabet{\mathsfit}{T1}{\sfdefault}{\mddefault}{\sldefault}
\SetMathAlphabet{\mathsfit}{bold}{T1}{\sfdefault}{\bfdefault}{\sldefault}

\newcommand*{\Scale}[2][4]{\scalebox{#1}{$#2$}} 
\newcommand{\romansubs}{\renewcommand{\theequation}{\theparentequation \roman{equation}}} 
\newcommand{\Eone}{\mathsfit{E}_{1}}
\newcommand{\Etwo}{\mathsfit{E}_{2}}
\newcommand{\Ethree}{\mathsfit{E}_{3}}

\newcommand{\PRLsep}{\noindent\makebox[\linewidth]{\resizebox{0.750\linewidth}{1pt}{$\blacklozenge$}}\bigskip}

\begin{document}
\pagestyle{fancy}
\fancyhead{} 
\fancyhead[OR]{\thepage}
\fancyhead[OC]{{\small{
   \textsf{Non-com\-mut\-a\-tive black holes}}}}
\fancyfoot{} 
\renewcommand\headrulewidth{0.5pt}
\addtolength{\headheight}{2pt} 
\global\long\def\tdud#1#2#3#4#5{#1_{#2}{}^{#3}{}_{#4}{}^{#5}}
\global\long\def\tudu#1#2#3#4#5{#1^{#2}{}_{#3}{}^{#4}{}_{#5}}

\twocolumn

\title{\vspace{-2cm}\hspace{-0.0cm}\rule{\linewidth}{0.2mm}\\
\bf{\Large{\textsf{Non-com\-mut\-a\-tive black holes of various genera in the connection formalism}}}}


\author[1]{\small{Mathew Schneider}\thanks{\href{mailto:mlschnei@sfu.ca}{mlschnei@sfu.ca}}}%

\author[2,3]{\small{Andrew DeBenedictis}\thanks{
   \href{mailto:adebened@sfu.ca}{adebened@sfu.ca}}}%


\affil[1]{
   \footnotesize{\it{Department of Physics, Simon Fraser University}}\\
   \footnotesize{\it{8888 University Drive, Burnaby, BC, V5A 1S6, Canada}} \protect \\
   \footnotesize{ . }
}

\affil[2]{\footnotesize{\it{The Pacific Institute
   for the Mathematical Sciences}} \protect\\
   \footnotesize{and}
}

\affil[3]{\footnotesize{\it{Department of Physics, Simon Fraser University}}\\
   \footnotesize{\it{8888 University Drive, Burnaby, BC, V5A 1S6, Canada}}
}

\date{\vspace{-0.8cm}({\footnotesize{June 23, 2020}})} 
\twocolumn[ 
  \begin{@twocolumnfalse}  
  \begin{changemargin}{1.75cm}{1.75cm} 
\maketitle
\end{changemargin}
\vspace{-1.0cm}
\begin{changemargin}{1.5cm}{1.5cm} 
\begin{abstract}
{\noindent\small{We consider black hole interiors of arbitrary genus number within the paradigm of non-com\-mut\-a\-tive geometry. The study is performed in two ways: One way is a simple smearing of a matter distribution within the black hole. The resulting structure is often known in the literature as a ``model inspired by non-com\-mut\-a\-tive geometry''. The second method involves a more fundamental approach, in which the Hamiltonian formalism is utilized and a non-trivial Poisson bracket is introduced between the configuration degrees of freedom, as well as between the canonical momentum degrees of freedom. This is done in terms of connection variables instead of the more common ADM variables. Connection variables are utilized here since non-com\-mut\-a\-tive effects are usually inspired from the quantum theory, and it is the connection variables that are used in some of the more promising modern theories of quantum gravity. We find that in the first study, the singularity of the black holes can easily be removed. In the second study, we find that introducing a non-trivial bracket between the connections (the configuration variables) may delay the singularity, but not necessarily eliminate it. However, by introducing a non-trivial bracket between the densitized triads (the canonical momentum variables) the singularity can generally be removed. In some cases, new horizons also appear due to the non-com\-mut\-a\-tiv\-i\-ty.}}
\end{abstract}
\noindent{\footnotesize PACS(2010): 11.10.Nx \; 02.40.Gh \; 45.20.Jj \; 47.10.Df}
\rule{\linewidth}{0.2mm}
\end{changemargin}
\end{@twocolumnfalse} 
]
\saythanks 
\vspace{0.5cm}
{\setstretch{0.9} 
\section{Introduction}
The general theory of relativity has, to date, robustly passed a number of experimental tests. These tests are no longer limited to the arena of weak-field gravity but also, due to more recent gravitational wave detection events, strong field regimes such as black hole mergers \cite{ref:ligostuff}. As successful as general relativity is, there should be some way to reconcile the fundamental properties of matter fields sourcing gravity (which at the fundamental level are quantum in nature) with the gravitational field that the matter produces. This compatibility could come from a theory of quantum gravity. General relativity, however, possesses the fundamental symmetry of background independence, and this makes the theory difficult to quantize in traditional manners \cite{ref:structural}, \cite{ref:bckgndindepstatusrept}. At the moment there are a number of candidate theories of quantum gravity which are in various stages of development \cite{ref:oritibook}-\cite{ref:ideasQG} although none can yet be seen as a complete theory of quantum gravitation. Because of this it is useful at the classical level to attempt to glean what some effects of a quantum theory of gravity may be. 

One issue that is believed to be resolved in a quantum gravity theory is that of the gravitational singularities predicted by various classical theories of gravity. The most famous of these singularities reside in the realm of early universe cosmology, and black hole interiors. It is the latter issue that we wish to discuss in this paper.

The fundamental mathematical object on which quantum theory is based on is the non-trivial commutator between a system's configuration variables and associated canonical momentum variables. At the level of classical mechanics this manifests itself as a non-trivial Poisson bracket. The field of non-com\-mut\-a\-tive geometry augments this structure by introducing, as well as the usual bracket between configuration-momentum variables, a non-trivial bracket between configuration variables. At the level of usual quantum mechanics this would be a non-trivial commutator of the form
\begin{equation}
 \left[x^{a},\,y^{b}\right]= i\epsilon^{ab}_{\;\;\;c}\theta^{c}\, \label{eq:noncommutatorx}
\end{equation}
where $\theta^{c}$ is a vector whose entries measure the amount of non-com\-mut\-a\-tiv\-ity between the various coordinates. The bracket (\ref{eq:noncommutatorx}) of course implies an uncertainty relation between different coordinates, and sets a limit on the amount of loc\-al\-iz\-a\-tion a particle may have. A measurement along one axis to high precision comes at the expense of losing some information along another axis. Hence, geometry in this sense really does become non-com\-mut\-a\-tive. 

It is natural to then further extend the theory to include a non-trivial bracket between the canonical momenta as
\begin{equation}
 \left[p_{a},\,p_{b}\right]= i\epsilon_{ab}^{\;\;\;c}\beta_{c}\, \label{eq:noncommutatorp}
\end{equation}
leading to a similar uncertainty between the measurement of momenta in different principal directions. 

Non-com\-mut\-a\-tive quantum theories have been studied in various fields of physics. The original paper seems to be the pioneering work of Hartland Snyder \cite{ref:snyder} and since then there has been much application of non-com\-mut\-a\-tive geometry to theoretical physics. (See \cite{ref:ncgphysstart} - \cite{ref:ncgphysend} and references therein.)

At the classical level the new commutators should manifest themselves as an extension of the usual Poisson algebra of ordinary classical mechanics, leading to a type of ``non-com\-mut\-a\-tive classical mechanics''. In non-com\-mut\-a\-tive mechanics the usual Poisson algebra is deformed via the introduction of a deformed Moyal product. That is, the brackets of non-com\-mut\-a\-tive mechanics are calculated via
\begin{equation}
 \left\{f,\, g\right\}:= f\star g -g\star f\, ,\label{eq:moyalpoisson} 
\end{equation}
where the Moyal product here is defined as
\begin{equation}
 \left(f\star g\right) (v):=\exp\left[\frac{1}{2}w^{ab} \partial_{a}\tilde{\partial}_{b}\right]f(v)\,g(\tilde{v})_{|\tilde{v}=v}\,. \label{eq:explicitmoyal}
\end{equation}
Here operators with a tilde operate only on tilde coordinates and un-tilded operators operate on un-tilded coordinates. In the end, the two sets of coordinates are made coincident. The matrix $w^{ab}$ represents the deformed symplectic form
\begin{equation}
 w^{ab}=\left[
    \begin{array}{c;{2pt/2pt}c}
        \begin{matrix}
       \epsilon^{ab}_{\;\;\,c}\theta^{c}
        \end{matrix}  &
        \begin{matrix}
        -\delta^{ab} 
        \end{matrix} \\ \hdashline[2pt/2pt]
        \begin{matrix}
       \delta^{ab} 
        \end{matrix} &
         \begin{matrix}
        \epsilon^{ab}_{\;\;\,c}\beta^{c}
        \end{matrix}
    \end{array}
\right]\,. \label{eq:sympform}
\end{equation}
We note here that there is actually a further correction to the above symplectic form but it is proportional to the product $\theta^{a}\beta^{b}$ and hence we ignore it as both these parameters are assumed to be small \cite{ref:djemai}. It may be seen by explicit calculation that in the limit $\theta^{c}=0=\beta^{c}$ the expression in (\ref{eq:explicitmoyal}) yields the usual Poisson brackets of ordinary classical mechanics. Explicitly, (\ref{eq:moyalpoisson}) and (\ref{eq:explicitmoyal}), using (\ref{eq:sympform}) yields 
\begin{subequations}
\romansubs
{\allowdisplaybreaks\begin{align}
 \left\{p_{a},\, x^{b}\right\} = & \, \delta_{a}^{b}\,,\\
 \left\{x^{a},\, x^{b}\right\} = & \, \epsilon^{ab}_{\;\;\;c}\theta^{c} \,,\\
 \left\{p_{a},\, p_{b}\right\} = & \, \epsilon_{ab}^{\;\;\;c}\beta_{c}\,.
\end{align}}
\end{subequations}

Reviews of non-com\-mut\-a\-tive mechanics may be found in \cite{ref:djemai} and \cite{ref:gouba} and references therein.

The transition from particle mechanics to field theories is not necessarily straight forward, particularly in the realm of gravitation \cite{ref:ncgeinststart}-\cite{ref:ncgeinstend}. However, if one symmetry reduces the system to minisuperspace models, then it can be argued that one augments the field Poisson algebra in a similar manner to what is done in the particle mechanics \cite{ref:minisupstart}-\cite{ref:minisupend}. That is, in a minisuperspace model with fields $\psi^{a}$ and corresponding canonical momenta $\pi_{b}$ we have
\begin{subequations}
\romansubs
{\allowdisplaybreaks\begin{align}
 \left\{\pi_{a}(x),\, \psi^{b}(y)\right\} \propto & \, \delta_{a}^{b}\delta(x-y)\,,  \label{eq:fieldpoisson1}\\
 \left\{\psi^{a}(x),\, \psi^{b}(y)\right\} \propto & \, \epsilon^{ab}_{\;\;\;c}\theta^{c}\delta(x-y)  \,, \label{eq:fieldpoisson2}\\
 \left\{\pi_{a}(x),\, \pi_{b}(y)\right\} \propto & \, \epsilon_{ab}^{\;\;\;c}\beta_{c}\delta(x-y)\,. \label{eq:fieldpoisson3}
\end{align}}
\end{subequations}

As the brackets are modified from the canonical ones, it is possible that such an algebraic deformation introduces an anomaly in the gravitational constraint algebra. In the symmetry-frozen homogeneous scenarios variable deformations generally do not introduce such anomalies as the algebra trivializes due to the vanishing of spatial derivatives, and being able to globally set the shift vector to zero. The situation likely needs further study under algebraic deformations, but it is generally believed that at high energies, non-com\-mut\-a\-tive effects would anyway alter the symmetry of the low energy theory \cite{ref:anomaly}, \cite{ref:moia}, so it is not clear if one should demand low energy symmetries to hold in the regime where non-com\-mut\-a\-tive effects become important. Still, one needs to be cautious in interpreting results in such potentially symmetry-broken theories. The general issue for the case of {\emph{variable}} deformations is summarized in \cite{ref:tmma}, and the situation with non-com\-mut\-ing coordinates, and their generalized non-com\-mut\-a\-tiv\-i\-ty via the Seiberg-Witten map technique and including generalizations to the Poincar\'{e} gauge gravity approach was studied in \cite{ref:mukherjee2}.

It is also interesting to note that aspects of spin can be viewed in the paradigm of non-com\-mut\-a\-tive geometry \cite{ref:deriglazov1}, including within general relativity \cite{ref:deriglazov2}. 

This manuscript is laid out as follows. In section \ref{sec:smearing} we analyze models where the the black holes are supported by a smeared out distribution of material, which is sometimes performed in the literature as an approximation of non-com\-mut\-a\-tive effects on the matter fields due to the non-loc\-al\-iz\-a\-tion that non-com\-mut\-a\-tive geometry introduces. In section \ref{sec:hamiltonian} the non-com\-mut\-a\-tiv\-ity is manifestly included in the brackets of the Poisson algebra in the configuration and momentum variables of the gravitational Hamiltonian system. The study there is performed in the connection formalism as this formalism is seen as a promising avenue to a theory of quantum gravity. Finally we conclude with a brief summary of the findings.

\section{Smearing of the matter distribution} \label{sec:smearing}
The method used here is often said to be ``inspired by non-com\-mut\-a\-tive geometry''. The idea here is quite simple and straight-forward and mainly serves as a segue to the Hamiltonian analysis of the next section. For concreteness in setting up the problem and method, we will assume at the moment that the black hole is a spherically symmetric one, but the ideas apply to all the types of metrics considered in this work. Consider the Einstein equations in mixed form
\begin{equation}
 R^{\mu}_{\;\,\nu}-\frac{1}{2}R\delta^{\mu}_{\;\,\nu}=8\pi T^{\mu}_{\;\,\nu}\, . \label{eq:einsteq}
\end{equation}
If one restricts these equations to spherical symmetry by utilizing the following line element
\begin{align}
 ds^{2}=&-\exp\left(\alpha(r,t)\right)\,dt^{2} + \exp\left(\beta(r,t)\right)\, dr^{2} \nonumber \\
 & + r^{2}\,d\theta^{2} +r^{2}\sin^{2}\theta\, d\phi^{2}\, , \label{eq:metric}
\end{align}
then equations (\ref{eq:einsteq}) may be manipulated to yield the following solution, assuming the stress-energy tensor components $T^{t}_{\;\,t}$ and $T^{r}_{\;\,r}$ are free parameters \cite{ref:syngebook}\footnote{We cannot prescribe more than two functions, as Einstein's spherically symmetric equations with matter are under determined by precisely two \cite{ref:dasdebbook}.} \cite{ref:novaarticle}:
\begin{subequations}
\romansubs
{\allowdisplaybreaks\begin{align}
e^{-\beta}=&1+\frac{8\pi}{r} \left[\int_{0_{+}}^{r}
T^{t}_{\;t}(x,t)\,x^{2}\,dx\right], \label{eq:ebeta} \\
e^{\alpha}=&e^{-\beta} \left\{\exp\left[h(t)+8\pi\int_{0_{+}}^{r}
\left[T^{r}_{\;r}(x,t) \right.\right.\right. \nonumber \\
&-\left.\left.\left.T^{t}_{\;t}(x,t)\right] e^{\alpha(x,t)}
x\,dx\right]\right\}, \label{eq:ealpha}\\
\Scale[0.95]{T^{r}_{\;t}:=}&\Scale[0.95]{\frac{1}{r^{2}} \left[ 2f(t)\dot{f}(t)
-\int_{0_{+}}^{r}T^{t}_{\;t,t}(x,t) x^{2}\,dx\right],} \label{eq:energyflux} \\
T^{\theta}_{\;\theta}\equiv T^{\phi}_{\;\phi}
:=&\frac{r}{2}\left[T^{r}_{\;r,r}+T^{t}_{\;r,t}\right]
+\left[1+\frac{r}{4}\alpha_{,r}\right]T^{r}_{\;r} \nonumber \\
& +\frac{r}{4}\left(\alpha+\beta\right)_{,t}T^{t}_{\,r}
-\frac{r}{4}\alpha_{,r}T^{t}_{\;t}\,, \label{eq:transpressure}
\end{align}}
\end{subequations}
where a comma denotes partial differentiation. Equation (\ref{eq:energyflux}) is defined from the $r-t$ Einstein equation, and (\ref{eq:transpressure}) is defined from the conservation law. Now, the Schwarzschild metric may be seen as a solution to the above equations with a ``point mass'' located at $r=0$. That is, one may prescribe
\begin{equation}
T^{t}_{\;t}(r)= -\frac{M}{4\pi r^{2}}\, \delta(r), \;\;\; T^{r}_{\;r}=0\,. \label{eq:schwsettens}
\end{equation}
It is straight-forward, by inserting (\ref{eq:schwsettens}) into equations (\ref{eq:ebeta})-(\ref{eq:ealpha}), to see that the resulting metric functions, $e^{\alpha}$ and $e^{\beta}$ yield, after a trivial re-scaling of the $t$ coordinate, the famous Schwarzschild metric
\begin{equation}
 e^{\alpha}=\left(1-\frac{2M}{r}\right) = e^{-\beta}\,. \nonumber
\end{equation}

In non-com\-mut\-a\-tive geometry inspired models, one smears the matter distribution (\ref{eq:schwsettens}) on a scale proportional to the coordinate non-com\-mut\-a\-tiv\-ity parameter, $\theta$. The argument is that the matter is not completely localized due to the uncertainty principle between coordinates brought on by the non-com\-mut\-a\-tiv\-ity. Such inspired models have been studied in \cite{ref:inspiredbhstart}-\cite{ref:inspiredbhend} for spherical black holes without cosmological constant, and in \cite{ref:noncomkerr}, \cite{ref:modesto} for rotating black holes. In \cite{ref:rabin1} the relationship between inspired theories and the Voros product (instead of the Moyal one) has been explored. Studies of inspired models have been performed in \cite{ref:inspwhstart}-\cite{ref:inspwhend} with respect to wormholes.

We wish to extend the study here to encompass black holes beyond spherical, both in shape and in topology. This is done for the reason of consistency. That is, one wishes to study if and how singularities are affected in as many scenarios as possible to determine how universal the non-com\-mut\-a\-tive effects are. One can then make more general statements about non-com\-mut\-a\-tiv\-ity. We also include a cosmological constant, since in four dimensions a cosmological constant is required for black holes of exotic topology \cite{ref:topostart} - \cite{ref:topoend}.

As we are interested specifically in the singularity issue of black holes, we will be concentrating on the interior region. First we wish to re-write the line element (\ref{eq:metric}) in a form more appropriate for the study of black hole interiors and various topologies. The form is as follows:
\begin{align}
ds^{2}=&-e^{A(\tau)}\,d\tau^{2} + e^{B(\tau)}\,dy^{2} +\tau^{2}\,d\varrho^{2}  \nonumber \\
& \; + \tau^{2} c_{0} \,\sinh^{2}(\sqrt{d_{0}}\varrho)\,d\varphi^{2}\, , \label{eq:intmetric}
\end{align}
and reflects the fact that the interior region is time dependent and that the exterior radial coordinate, $r$, is timelike in the interior region (we do not consider cases here with inner horizons)\footnote{In the coordinate chart of (\ref{eq:intmetric}) the familiar Schwarzschild line element takes the form $ds^{2}=-\frac{d\tau^{2}}{\frac{2M}{\tau}-1}+\left(\frac{2M}{\tau}-1\right)\,dy^{2}+\tau^{2}\,d\varrho^{2} +\tau^{2}\sin^{2}\hspace{-0.05cm}\varrho\,d\varphi^{2}$ with $\tau < 2M$.}. We are considering time dependence only, due to the fact that we are smearing classical non-rotating \emph{vacuum} black holes, save for the ``point'' source, whose corresponding interiors are also homogeneous. The constants $c_{0}$ and $d_{0}$ dictate the compatible topology of the spacetime's two-dimensional subspaces. The various cases are as follows:\\[0.2cm]
i) $d_{0}=-1$, $c_{0}=-1$: In this scenario $(\varrho,\,\phi)$ sub-manifolds are spheres. \\
ii) $d_{0}=0$, $\underset{d_{0}\rightarrow 0}\lim\,c_{0}=\frac{1}{d_{0}}$: In this scenario $(\varrho,\,\phi)$ sub-manifolds are tori (and the sub-manifolds for this case are intrinsically flat). \\
iii) $d_{0}=1$, $c_{0}=1$: In this case $(\varrho,\,\phi)$ sub-manifolds are surfaces of constant negative curvature of genus $g > 1$, depending on the identifications chosen. Such surfaces may be compact or not \cite{ref:topoend}, \cite{ref:nakahara}. \\[0.2cm]
In the spherical case, such solutions are sometimes referred to in the literature as ``T-spheres'' \cite{ref:tspherestart} - \cite{ref:tsphereend} and the time dependent domain inside the event horizon is sometimes referred to as the ``T-domain'' of the black hole.

Einstein's equations, for the line element (\ref{eq:intmetric}) yield the following general solution analogous to (\ref{eq:ebeta})-(\ref{eq:transpressure}), assuming here time dependence only:\\[0.2cm]

\begin{subequations}
\romansubs
{\allowdisplaybreaks\begin{align}
e^{-A}=&d_{0}-\frac{8\pi}{\tau} \left[\int_{\tau_{1}}^{\tau}
T^{y}_{\;y}(\tau^{\prime})\,\tau^{\prime\,2}\,d\tau^{\prime}\right], \label{eq:eA} \\
e^{B}=&e^{-A} \left\{\exp\left[k_{0}-8\pi \int_{\tau_{1}}^{\tau}
\left[T^{\tau}_{\;\tau}(\tau^{\prime}) \right.\right.\right. \nonumber \\
&-\left.\left.\left.T^{y}_{\;y}(\tau^{\prime})\right] e^{A(\tau^{\prime})}
\tau^{\prime}\,d\tau^{\prime}\right]\right\}, \label{eq:eB}\\
T^{\tau}_{\;y}=& 0\,, \label{eq:intenergyflux} \\
T^{\varrho}_{\;\varrho}\equiv T^{\varphi}_{\;\varphi}
:=&\frac{\tau}{2}T^{\tau}_{\;\tau,\tau}
+\left[1+\frac{\tau}{4}B_{,\tau}\right]T^{\tau}_{\;\tau} \nonumber \\
&- \frac{\tau}{4}B_{,\tau}T^{y}_{\;y}\,, \label{eq:inttranspressure}
\end{align}}
\end{subequations}

In the case of of a ``point'' source (in the interior region $T^{y}_{\;y\,\mbox{\tiny{matter}}}(\tau)= -\frac{M}{4\pi \tau^{2}}\, \delta(\tau)$), supplemented with cosmological constant $\left(T^{\tau}_{\;\tau\,\mbox{\tiny{$\Lambda$}}}=T^{y}_{\;y\,\mbox{\tiny{$\Lambda$}}}=-\Lambda/(8\pi)\right)$, the above solutions yield, after a rescaling of the $y$ coordinate:
\begin{equation}
 e^{-A(\tau)}=\left(d_{0}+\frac{2M}{\tau}+\frac{\Lambda}{3}\tau^{2}\right)=e^{B(\tau)}\,. \label{eq:classicmetric}
\end{equation}
Such black hole solutions have been studied in detail in \cite{ref:topostart} - \cite{ref:topoend}, and within quantum gravity theories in \cite{ref:ourent} - \cite{ref:ourwdw}.

The non-com\-mut\-a\-tive smearing is often performed via the implementation of replacing the ``point'' source with a Gaussian or Lorentzian whose characteristic width is of the scale of the non-com\-mut\-a\-tiv\-ity parameter, $\theta$. Without guidance from experiment, this is usually taken to be of the order of the Planck length. We consider here the following profile curves for $T^{y}_{\;y\,\mbox{\tiny{matter}}}(\tau)$:

\begin{subequations}
\romansubs
{\allowdisplaybreaks\begin{align}
T^{y}_{\;y\,\mbox{\tiny{matter}}}(\tau) =&  - \frac{M\sqrt{\theta}}{\pi^2(\tau^2 + \theta)^2} \,, \label{eq:Lorentzian} \\
T^{y}_{\;y\,\mbox{\tiny{matter}}}(\tau) =& -\frac{M}{(4\pi\theta)^{3/2}} e^{-\frac{\tau^2}{4\theta}}  \,, \label{eq:Gaussian}
\end{align}}
\end{subequations}
the first profile being Lorentzian and the second Gaussian. Further, since the matter profile no longer vanishes abruptly, we relate the above stress-energy components to their corresponding local energy densities via an equation of state, for which we take the polytropic form. That is, in both the Lorentzian and Gaussian scenarios we prescribe an energy density via
\begin{equation}
 T^{\tau}_{\;\tau\,\mbox{\tiny{matter}}}(\tau)=k\, \left(T^{y}_{\;y\,\mbox{\tiny{matter}}}(\tau)\right)^{\gamma} \, , \label{eq:polytropic}
\end{equation}
where $k$ and $\gamma$ are constants. This form works well in idealized studies of stellar structure \cite{ref:polystart} - \cite{ref:polyend} and seems a natural choice for the type of exotic ``star'' we are studying here.

By using (\ref{eq:Lorentzian}) and (\ref{eq:Gaussian}) in equations (\ref{eq:eA}) and (\ref{eq:eB}) one arrives at the following analytical solutions:
\begin{equation}
\Scale[.93]{e^{-A(\tau)} = d_0  + \frac{\Lambda}{3}\tau^2 + \frac{6M}{\pi\tau}\arctan(\tau/\sqrt{\theta}) - \frac{6M\sqrt{\theta}}{\pi(\tau^2 + \theta)}}
\end{equation}
for the Lorentzian case and 
\begin{equation}
\Scale[.95]{e^{-A(\tau)} = d_0  + \frac{\Lambda}{3}\tau^2 + \frac{2M}{\tau}\text{erf}(\tau/2\sqrt{\theta}) - \frac{2M}{\sqrt{\pi\theta}}e^{-\frac{\tau^2}{4\theta}}}
\end{equation}
for the Gaussian scenario. Solutions for $e^{B(\tau)}$ were also obtained, but being rather complicated and in terms of quadrature, are not displayed here.

Of particular interest here is to study the properties of what replaces the singularity of the com\-mut\-a\-tive theory. To facilitate this we calculate the components of the Riemann curvature tensor in an orthonormal (hatted) frame,
\begin{equation}
 R_{\hat{\mu}\hat{\nu}\hat{\rho}\hat{\sigma}}=R_{\alpha\beta\gamma\delta}h_{\hat{\mu}}^{\;\,\alpha}h_{\hat{\nu}}^{\;\,\beta}h_{\hat{\rho}}^{\;\,\gamma}h_{\hat{\sigma}}^{\;\,\delta}\,. \label{eq:orthriem}
\end{equation}
Here $h_{.}^{\;.}$ represent the components of a local orthonormal tetrad. We choose specifically the tetrad coincident with the coordinate directions. That is, the tetrad is given by
\begin{align}
 h_{\hat{\tau}}^{\;\mu}=&\frac{\delta_{\tau}^{\;\mu}}{e^{A(\tau)/2}},\; h_{\hat{y}}^{\;\mu}=\frac{\delta_{y}^{\;\mu}}{e^{B(\tau)/2}}\, \nonumber \\
 h_{\hat{\varrho}}^{\;\mu}=& \frac{\delta_{\varrho}^{\;\mu}}{\tau},\; h_{\hat{\varphi}}^{\;\mu}=\frac{\delta_{\varphi}^{\;\mu}}{\tau |\sqrt{c_{0}}\sinh\left(\sqrt{d_{0}}\varrho\right)|}\,. \nonumber
\end{align}

The resulting orthonormal Riemann components are rather lengthy and do not reveal much due to their complication. It is useful therefore to present the lowest order terms in a series expansion about the com\-mut\-a\-tive solution's singular point ($\tau=0$). Such an expansion yields the following components, plus those related by symmetries, for the Lorentzian case:
\begin{subequations}
\romansubs
{\allowdisplaybreaks\begin{align}
R_{\hat{\tau} \hat{y} \hat{\tau} \hat{y}}=& - \frac{\Lambda}{3} + 4k\pi\bigg(\frac{M}{\pi^2\theta^{3/2}}\bigg)^\gamma + \frac{4M}{3\pi\theta^{3/2}} \nonumber \\
&+ \mathcal{O}(\tau^{2})\,, \label{eq:RtytyL} \\
R_{\hat{\tau} \hat{\varrho} \hat{\tau} \hat{\varrho}}=& -\frac{\Lambda}{3} - \frac{8M}{3\pi\theta^{3/2}} + \frac{32M}{5\pi\theta^{5/2}}\tau^2 + \mathcal{O}(\tau^{4})\,, \label{eq:RtrhotrhoL} \\
R_{\hat{y} \hat{\varrho} \hat{y} \hat{\varrho}}=& ~\frac{\Lambda}{3} -4k\pi\bigg(\frac{M}{\pi^2\theta^{3/2}}\bigg)^\gamma - \frac{4M}{3\pi\theta^{3/2}} \nonumber \\
 &+ \frac{8k\pi\gamma}{\theta}\bigg(\frac{M}{\pi^2\theta^{3/2}}\bigg)^\gamma \tau^2 + \frac{8M}{5\pi\theta^{5/2}} \tau^2 \nonumber \\
 &+ \mathcal{O}(\tau^{4})\, \label{eq:RyrhoyrhoL} \\
R_{\hat{\varrho} \hat{\varphi}  \hat{\varrho} \hat{\varphi} }=& ~\frac{\Lambda}{3} + \frac{8M}{3\pi\theta^{3/2}} - \frac{16M}{5\pi\theta^{5/2}}\tau^2 + \mathcal{O}(\tau^{4})\, \label{eq:RrhophirhophiL}\,, 
\end{align}}
\end{subequations}
and for the Gaussian case:
\begin{subequations}
\romansubs
{\allowdisplaybreaks\begin{align}
R_{\hat{\tau} \hat{y} \tau \hat{y}}=&  - \frac{\Lambda}{3} + 2k\pi\bigg(\frac{M}{(4\pi\theta)^{3/2}}\bigg)^\gamma +  \frac{M}{12\sqrt{\pi}\theta^{3/2}} \nonumber \\
&+ \mathcal{O}(\tau^{2})\,, \label{eq:RtytyG} \\
R_{\hat{\tau} \hat{\varrho} \hat{\tau} \hat{\varrho}}=& ~\frac{\Lambda}{3} - \frac{M}{3\pi\theta^{3/2}} + \frac{M}{10\sqrt{\pi}\theta^{5/2}}\tau^2 + \mathcal{O}(\tau^{4})\,, \label{eq:RtrhotrhoG} \\
R_{\hat{y} \hat{\varrho} \hat{y} \hat{\varrho}}=&  ~\frac{\Lambda}{3} - 4k\pi\bigg(\frac{M}{(4\pi\theta)^{3/2}}\bigg)^\gamma -  \frac{M}{6\sqrt{\pi}\theta^{3/2}} \nonumber \\
 & + \frac{k\pi\gamma}{\theta}\bigg(\frac{M}{(4\pi\theta)^{3/2}}\bigg)^\gamma \tau^2 + \frac{M}{40\sqrt{\pi}\theta^{5/2}} \nonumber \\
 &+ \mathcal{O}(\tau^{4})\, \label{eq:RyrhoyrhoG} \\
R_{\hat{\varrho} \hat{\varphi}  \hat{\varrho} \hat{\varphi} }=& ~\frac{\Lambda}{3} + \frac{M}{3\pi\theta^{3/2}} - \frac{M}{20\sqrt{\pi}\theta^{5/2}}\tau^2+ \mathcal{O}(\tau^{4})\,. \label{eq:RrhophirhophiG}
\end{align}}
\end{subequations}
The results above are ``universal'' at $\tau=0$ in the sense that the topological parameter, $d_{0}$, does not contribute at zeroth order. This parameter comes in at order $\tau^{2}$ in $R_{\hat{\tau} \hat{y} \tau \hat{y}}$ (although this term is not shown due to its length), and at order $\tau^{4}$ or higher in the other components.

It may be noted that none of the components in (\ref{eq:RtytyL})-(\ref{eq:RrhophirhophiG}) are singular for finite $\theta$ and therefore the classical singularity present in the com\-mut\-a\-tive theory is removed. We should point out here that this result is not really surprising. One has excised the singular distribution of (\ref{eq:schwsettens}) and replaced it with a smooth distribution. In the T-domain this forced smearing is accompanied by the  expected energy condition violations which circumvent the singularity theorems for such spacetimes. Therefore, at the level of non-com\-mut\-a\-tive geometry inspired models of black holes, the non-com\-mut\-a\-tiv\-ity introduces energy condition violation on scales set by the non-com\-mut\-a\-tiv\-ity parameter $\theta$.  

We proceed next to a more rigorous analysis where the non-com\-mut\-a\-tiv\-ity is truly manifest in the algebra of the field variables.

\section{Hamiltonian evolution of black hole interiors}\label{sec:hamiltonian}
In this section we shall study the effects of non-com\-mut\-a\-tiv\-ity by directly supplementing the usual Poisson algebra of the gravitational Hamiltonian system with the additional structure on the configuration and momentum variables as in (\ref{eq:fieldpoisson1}-iii). We work here in the connection variables consisting of the $su(2)$ connection, which we denote $A^{i}_{\;a}$, and its conjugate momentum, the densitized triad, denoted by $E_{i}^{\;a}$ \footnote{Indices $i, j, k$ etc. denote $su(2)$ indices whereas indices $a,b,c$ etc. denote spatial indices.}.   These variables are chosen since they are the variables utilized in the theory of loop quantum gravity. At the quantum level, within the paradigm of non-com\-mut\-a\-tive geometry, the commutator between the configuration variables is taken to be non-trivial. As well, one may also take the commutator between the conjugate momenta as non-trivial. Working ``backwards'' towards the corresponding classical theory, these non-trivial commutators should manifest themselves as non-trivial Poisson brackets. 

It is generally accepted that loop quantum gravity puts forward a more promising approach towards a theory of quantum gravity than does the original Wheeler-DeWitt theory \cite{ref:structural}, \cite{ref:connection}, which utilizes ADM variables. Therefore, in light of this we choose to work in the variables of loop quantum gravity. With the modification of the Poisson brackets introduced by the extra non-com\-mut\-a\-tiv\-ity, it is possible that working in these variables is a different theory than working in the corresponding noncom\-mut\-a\-tive ADM theory. 

In terms of cosmological studies, a number of interesting noncom\-mut\-a\-tiv\-ity studies have been performed in \cite{ref:minisupstart}, \cite{ref:bastos},  \cite{ref:ncgcosmostart} - \cite{ref:ncgcosmoend} in standard variables.

\subsection{Black holes in connection variables}
Here we briefly review the mathematical structure of black holes in connection variables. In the connection variables which are utilized in the canonical formulation of loop quantum gravity, one begins with a $3+1$ decomposition of space-time where the metric is written in the usual way
\begin{equation}
 ds^{2}=\Scale[0.91]{-N^{2}\,d\tau^{2}+q_{ab}[dx^{a}+N^{a}\,d\tau][dx^{b}+N^{b}\,d\tau]\,,} \label{eq:admmetric}
\end{equation}
with $N$ the lapse and $N^{a}$ the shift vector. One then writes the resulting action in terms of the Ashtekar variables \cite{ref:ashvar}. These variables comprise of a generalized $su(2)$ connection $A^{i}_{\;a}$ and a densitized triad $E_{i}^{\;a}$. The connection field plays the role of the configuration variable, and is related to more familiar quantities as follows:
\begin{equation}
 A^{i}_{\;a}=\Gamma^{i}_{\;a} + \gamma K^{i}_{\;a}\,, \label{eq:ashcon}
\end{equation}
where $\Gamma^{i}_{\;a}$ is the ``fiducial'' spin connection
\begin{equation}
 \Gamma^{i}_{\;a}:=\epsilon^{ij}_{\;\;\;k}h_{j}^{\;b}\left(\partial_{[a}h^{k}_{\;b]}+\delta^{k\ell}\delta_{pq}h_{\ell}^{\;c}h^{p}_{\;a}\partial_{b}h^{q}_{\;c}\right)\,, \label{eq:spincon}
\end{equation}
and $K^{i}_{\;a}$ is the densitized extrinsic curvature
\begin{equation}
 K^{i}_{\;a}:= \frac{1}{\sqrt{\mbox{det}(E)}}K_{ab}E_{j}^{\;b}\delta^{ij}\,, \label{eq:extcurv}
\end{equation}
with $K_{ab}$ the usual extrinsic curvature of a $\tau =\mbox{const.}$ surface. The quantity $\gamma$ is known as the Barbero-Immirzi parameter. In the classical theory its value is arbitrary (though non-zero) but in the resulting quantum theory of loop quantum gravity, it must be set somehow. This is usually done by calculations of black hole entropy within the paradigm of loop quantum gravity and setting the result to one-quarter the area of the black hole  \cite{ref:lqgentstart}-\cite{ref:lqgentend}.

The momentum variable, $E_{i}^{\;a}$ is related to the three-metric, $q_{ab}\,$, via  
\begin{equation}
\mbox{det}(q)\,q^{ab}=E_{i}^{\;a}E_{j}^{\;b}\delta^{ij}\;. \label{eq:mettriadreln}
\end{equation}

These two variables, $A^{i}_{\;a}$ and $E_{i}^{\;a}$, are then the configuration and momentum variables respectively of the theory, subject to the Poisson algebra
\begin{equation}
 \left\{A^{i}_{\;a}(x),\,E_{j}^{\;b}(y)\right\}=8\pi \gamma \delta^{i}_{\;j} \delta_{a}^{\;b}\delta(x,\,y)\,, \label{eq:aebracket}
\end{equation}
with other brackets equal to zero. 

Via variation of the action with respect to the lapse and shift one obtains the Hamiltonian ($S$) and diffeomorphism constraints ($V_{b}$):
\begin{subequations}
\romansubs
{\allowdisplaybreaks\begin{align}
&\Scale[0.93]{S=\frac{E_{i}^{\;a}E_{j}^{\;b}}{\sqrt{\mbox{det}(E)}} \left[\epsilon^{ij}_{\;\;\;k}F^{k}_{\;ab}-2(1+\gamma^{2})K^{i}_{\;[a}K^{j}_{\;b]}\right]=0} \,,\label{eq:scalconst} \\
&V_{b}=E_{i}^{\;a}F^{i}_{\;ab}-(1+\gamma^{2})K^{i}_{\;b}G_{i}=0\,, \label{eq:vecconst}
\end{align}}
\end{subequations}
with $F^{i}_{\;ab}:=\partial_{a}A^{i}_{\;b}-\partial_{b}A^{i}_{\;a}+\epsilon^{i}_{\;jk}A^{j}_{\;a}A^{k}_{\;b}$\,. The extrinsic curvature quantities in (\ref{eq:scalconst}) are replaced with the connection via (\ref{eq:ashcon}) (the $h_{i}^{\;a}$ in $\Gamma^{i}_{\;a}$ being functions of $E_{i}^{\;a}$).

There is also the internal $SU(2)$ degree of freedom that can be fixed. (The metric, depending on the ``square'' of the densitized triad via (\ref{eq:mettriadreln}), allows for $SU(2)$ rotations which preserve the metric.) This gauge can be fixed via the Gauss constraint:
\begin{equation}
 G_{i}:=\partial_{a} E_{i}^{\;a} +\epsilon_{ij}^{\;\;\;k}A^{j}_{\;a}E_{k}^{\;a}=0\, . \label{eq:gaussconst}
\end{equation}

At this stage we need to choose an ansatz for the $su(2)$ connection and the densitized triad which is compatible with our geometries. An appropriate ansatz is provided by the following pair \cite{ref:ourent}, \cite{ref:oursu2}, \cite{ref:constcurve}:
\begin{subequations}
\romansubs
{\allowdisplaybreaks\begin{align}
A^{i}_{\;a}\uptau_{i}dx^{a}=&a_{3} \uptau_{y}\,dy + \left(a_{1} \uptau_{\varrho}+ a_{2} \uptau_{\varphi}\right)d\varrho \nonumber \\
&+\left( a_{2}\uptau_{\varrho} - a_{1}\uptau_{\varphi}\right)\sqrt{c_{0}}\sinh(\sqrt{d_{0}}\varrho)\,d\varphi \nonumber \\
& +\uptau_{y}\sqrt{c_{0}}\sqrt{d_{0}}\cosh(\sqrt{d_{0}}\varrho)\,d\varphi\,, \label{eq:hyperA} \\
{E}_{i}^{\;a}\uptau^{i}\partial_{a}=& -\Ethree \uptau_{y} \sqrt{c_{0}}\sinh(\sqrt{d_{0}}\varrho) \frac{\partial}{\partial y} \nonumber \\
&- \left(\Eone\uptau_{\varrho} +\Etwo\uptau_{\varphi}\right)\sqrt{c_{0}}\sinh(\sqrt{d_{0}}\varrho) \frac{\partial}{\partial \varrho} \nonumber \\
& +\left(\Eone\uptau_{\varphi}-\Etwo\uptau_{\varrho}\right) \frac{\partial}{\partial\varphi}\,, \label{eq:hyperE} 
\end{align}}
\end{subequations}
where $\uptau_{i}$ represent the $SU(2)$ generators. The functions $a_{I}$ and $\mathsfit{E}_{I}$ are functions of the interior time variable, $\tau$, only.

In terms of (\ref{eq:hyperE}), using (\ref{eq:mettriadreln}), a line-element of the form (\ref{eq:intmetric}) is written as
\begin{align}
 ds^{2}=&-N^{2}\,d\tau^{2}+\frac{(\Eone)^{2}+(\Etwo)^{2}}{\Ethree}dy^{2} +\Ethree\,d\varrho^{2} \nonumber \\
 &+\Ethree c_{0}\sinh^{2}(\sqrt{d_{0}}\varrho)\,d\varphi^{2}\,. \label{eq:triadline}
\end{align}

The Gauss constraint (\ref{eq:gaussconst}) for the cases considered here yields just one condition:
\begin{equation}
 a_{1}\Etwo-a_{2}\Eone = 0 \label{eq:gaugefixcond}
\end{equation}
which we will satisfy here by choosing
\begin{equation}
 a_{1}=0=\Eone\,. \label{eq:gaussfixing}
\end{equation}
The diffeomorphism constraint is automatically satisfied in these cases, leaving only the Hamiltonian (scalar) constraint. With the gauge fixing (\ref{eq:gaussfixing}), the resulting Hamiltonian constraint (supplemented with cosmological constant term) may be written as
\begin{align}
S=&-\frac{N\sqrt{c_{0}}}{2\sqrt{d_{0} \Ethree} \gamma^{2}}\left[\left((a_{2})^{2}-\gamma^{2}d_{0}\right)\Etwo+2\Ethree a_{2}a_{3}\right]\nonumber \\
&+\frac{N}{2\sqrt{\Ethree}\gamma^2}\Lambda \Etwo\Ethree\,.\label{eq:ourhamconst}
\end{align}
It should be noted that when integrating the Hamiltonian density the spatial variables have been integrated out and therefore the above result contains an (arbitrary) area from the $y$ and $\varrho$ integrals. This is set equal to one and we show below that this does not spoil the Hamiltonian evolution of the system. We also should add here that in constructing the field strength tensor, no further non-\-com\-mut\-a\-tiv\-ity was assumed beyond the usual com\-mut\-a\-tion relations amongst the $su(2)$ generators. The action here acquires no corrections as there are no Poisson brackets in its construction. A non-com\-mut\-a\-tiv\-ity would introduce a potential ambiguity if one wished to study the true quantum theory (not what is done here), where quantities in the Hamiltonian do not com\-mute. This would be analogous to the usual factor ordering ambiguity of going from a classical to a quantum theory.

At this stage one has all the ingredients required to study the evolution of the interior region of the black holes. The evolution proceeds according to the usual Hamiltonian equations of motion $\dot{a}_{2}=\left\{a_{2},\,S\right\}$, etc. subject to the usual Poisson algebra between the configuration variables, $a_{I}$, and their corresponding canonical momenta, $\mathsfit{E}_{I}$.

\subsection{Non-com\-mut\-a\-tive evolution of black holes}
Here we study the above gravitational system in connection variables, but where the usual Poisson algebra is augmented by the following brackets:
\begin{equation}
 \left\{a_{I},\,a_{J}\right\}=\epsilon_{IJ}\theta\,, \;\;\; \left\{\mathsfit{E}_{I},\,\mathsfit{E}_{J}\right\}=\epsilon_{IJ}\beta\,.
\end{equation}
We study scenarios where either $\theta$ or $\beta$ is zero, as well as those where neither parameter is zero. As this is a first study, we take the non-com\-mut\-a\-tive parameters, $\theta$ and $\beta$ to be constants. However, it is possible that they be modified in such a way that they depend on the metric properties (via the densitized triad) of the spacetime. This, for example, could arguably improve the theory by providing a natural way for the brackets to become less significant in low curvature regions.  

The resulting equations in the noncom\-mut\-a\-tive case are too complex to find analytic solutions so what we are solving here is a classic initial value problem. As such, initial conditions are required in order to study the evolution. We set initial conditions as follows: Note that the coordinate chart in use for the domain of (\ref{eq:triadline}) is $\tau < \tau_{\mbox{\tiny{H}}}$, where $\tau_{\mbox{\tiny{H}}}$ denotes the horizon value of $\tau$ and that the com\-mut\-a\-tive solution's singular point is located at $\tau=0$. The evolution is started far from the singular point, and relatively close to the horizon. We make the assumption that, far from the singular point, non-com\-mut\-a\-tive effects should be small as we know, for example, that the Schwarzschild solution is valid in moderately strong gravitational fields \cite{ref:exptschw}. (In fact com\-mut\-a\-tive general relativity seems to hold well even in the strong field regime \cite{ref:exptschw}, \cite{ref:grexpts}, so the noncom\-mut\-a\-tive results here really are expected to be manifest only when one is approaching the scale of quantum gravity effects.) Therefore on the initial time surface, which is far from the extremely strong field region, we set the values of of functions $a_{I}$ and $\mathsfit{E}_{I}$ set to their general relativity values. That is, the following initial values are used:
\begin{subequations}
\romansubs
{\allowdisplaybreaks\begin{align}
a_{3\mbox{\tiny{init}}}=-\frac{\gamma}{2N}e^{B}\dot{B}\,,\; & \; a_{2\mbox{\tiny{init}}}=-\frac{\gamma}{N}\,, \label{eq:ainits} \\
\mathsfit{E}_{3\mbox{\tiny{init}}}=\tau^{2}\,, \; & \; \mathsfit{E}_{2\mbox{\tiny{init}}}=e^{B/2}\tau\,, \label{eq:Einits}
\end{align}}
\end{subequations}
\noindent where the densitized triad components have been calculated via comparing (\ref{eq:triadline}) and (\ref{eq:intmetric}). The connection components have been calculated via a rather lengthy calculation utilizing (\ref{eq:ashcon}), (\ref{eq:spincon}) and (\ref{eq:extcurv}), using metric (\ref{eq:intmetric})'s triad pulled back to a $\tau=\mbox{const.}$ hypersurface. The function $B$ here is the com\-mut\-a\-tive solution's value given by (\ref{eq:classicmetric}). The lapse, $N$, is generally arbitrary but as we have set the coordinate system to be that of (\ref{eq:classicmetric})), we wish to use the time variable proportional to that used in (\ref{eq:classicmetric}). Therefore, we set the lapse equal to $\gamma^{2}\sqrt{\Ethree}/\Etwo$ and at this stage one may proceed with the evolution.

The resulting ``non-com\-mut\-a\-tive'' Hamilton equations of motion are given by:
\begin{subequations}
\romansubs
{\allowdisplaybreaks\begin{align}
\dot{a}_{2}=&\frac{\sqrt{c_{0}}\,\Ethree a_{2}a_{3}}{\sqrt{d_{0}}(\Etwo)^{2}} - \theta \frac{2\sqrt{c_{0}}a_{2}\Ethree}{\sqrt{d_{0}}\Etwo}\,, \label{eq:a2dot} \\
\dot{a}_{3}=-&\frac{2\sqrt{c_{0}}}{\sqrt{d_{0}}\,\Etwo}\left[a_{2}a_{3}-\frac{\Etwo}{2}\Lambda\right] \nonumber \\
&+ \theta \frac{\sqrt{c_{0}}}{\sqrt{d_{0}}\Etwo}\left[a_{2}\Etwo+a_{3}\Ethree\right]\,, \label{eq:a3dot} \\
\dot{\mathsfit{E}}_{2}=&\frac{\sqrt{c_{0}}}{\sqrt{d_{0}}\,\Etwo} \left[a_{2}\Etwo+a_{3}\Ethree\right] \nonumber \\
&-\beta \frac{2\sqrt{c_{0}}}{\sqrt{d_{0}}\Etwo} \left[a_{2}a_{3}-\frac{\Etwo}{2}\Lambda\right]\,, \label{eq:E2dot} \\
\dot{\mathsfit{E}}_{3}=&\frac{2\sqrt{c_{0}}\,a_{2}\Ethree}{\sqrt{d_{0}}\,\Etwo} -\beta \frac{\sqrt{c_{0}}\Ethree a_{2}a_{3}}{\sqrt{d_{0}}(\Etwo)^{2}}\,. \label{eq:E3dot}
\end{align}}
\end{subequations}
As mentioned previously, these equations are generally too complex to solve numerically, hence we illustrate the solutions below subject to the initial conditions provided by (\ref{eq:ainits}) and (\ref{eq:Einits}).

\subsubsection{Non-com\-mut\-a\-tive connection only}
Here we briefly summarize the results of the evolution of the above system subject to $\theta\neq 0$ and $\beta=0$. That is, here the standard theory is augmented with non-trivial configuration bracket only. For each of the three topological compatibilities (spherical, toroidal\-/cylindrical\-/planar, higher genus) the results are shown in the figures \ref{fig:thetasphds}-\ref{fig:thetahypds}.

\begin{figure}[h!t]
\centering
\vspace{0.0cm} \includegraphics[width=1.0\columnwidth, clip]{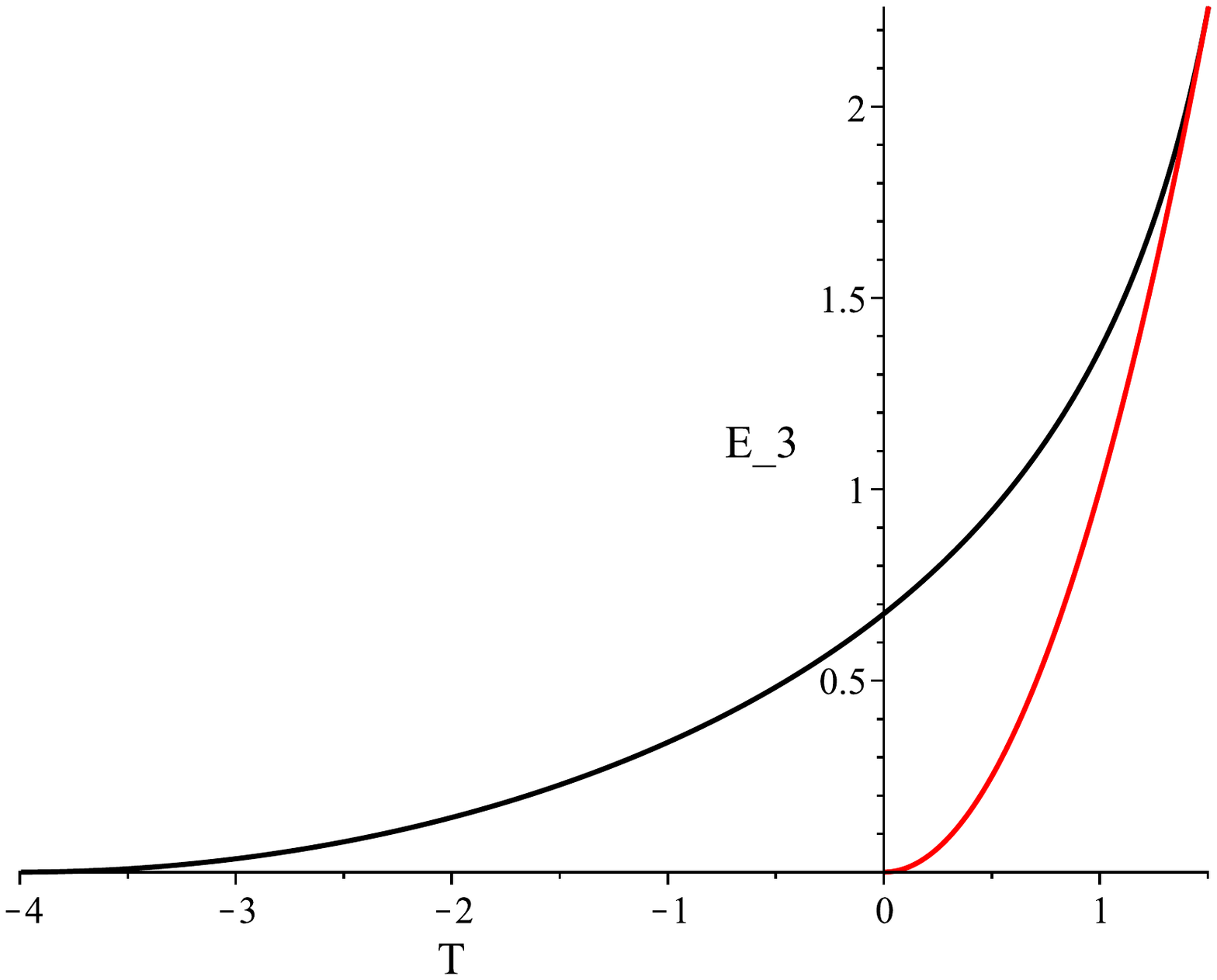}\\[-5.0cm]
\hspace{-0.3cm} \includegraphics[width=1.0\columnwidth, clip]{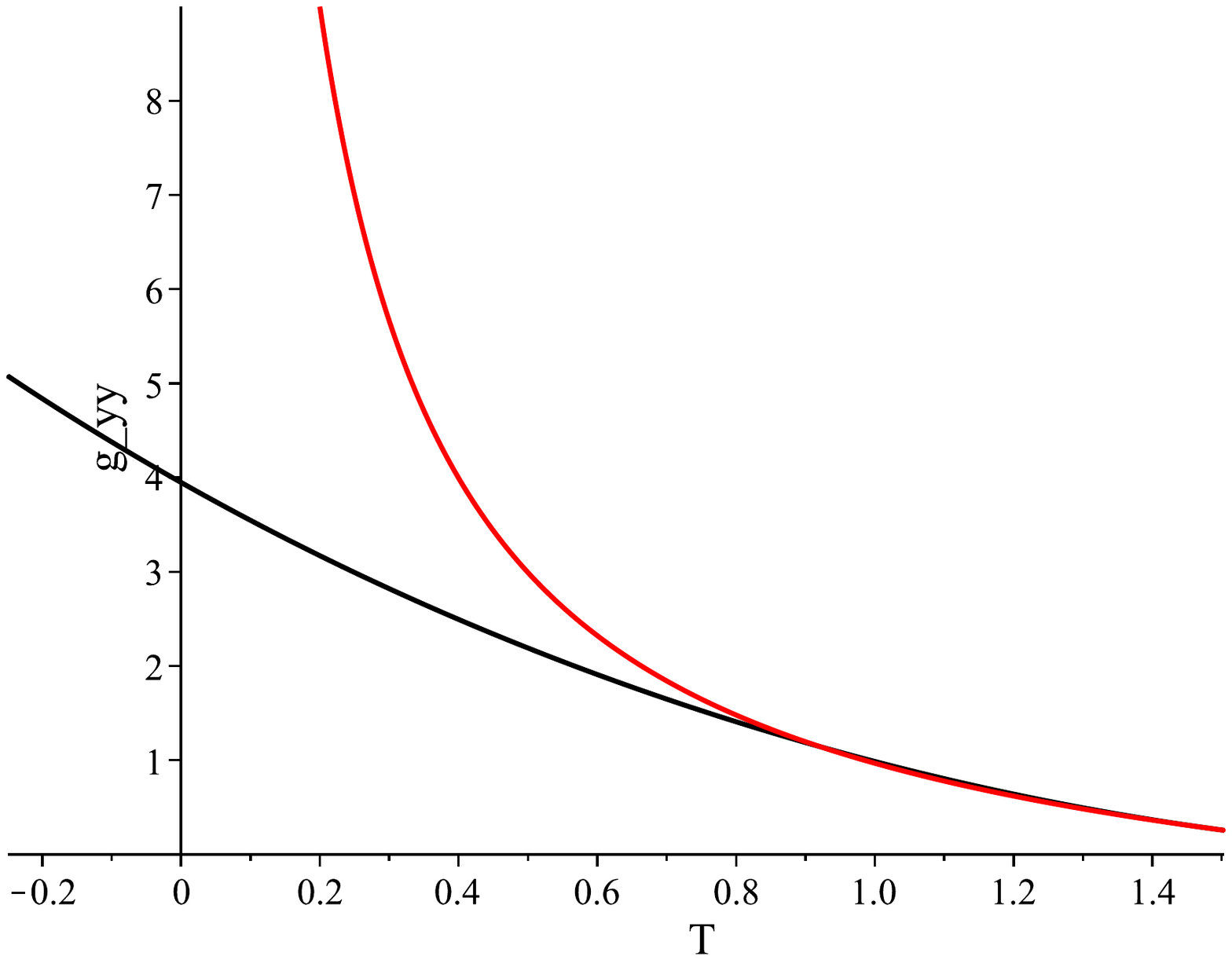}\\[-5.0cm]
\hspace{-0.3cm}\includegraphics[width=1.0\columnwidth, clip]{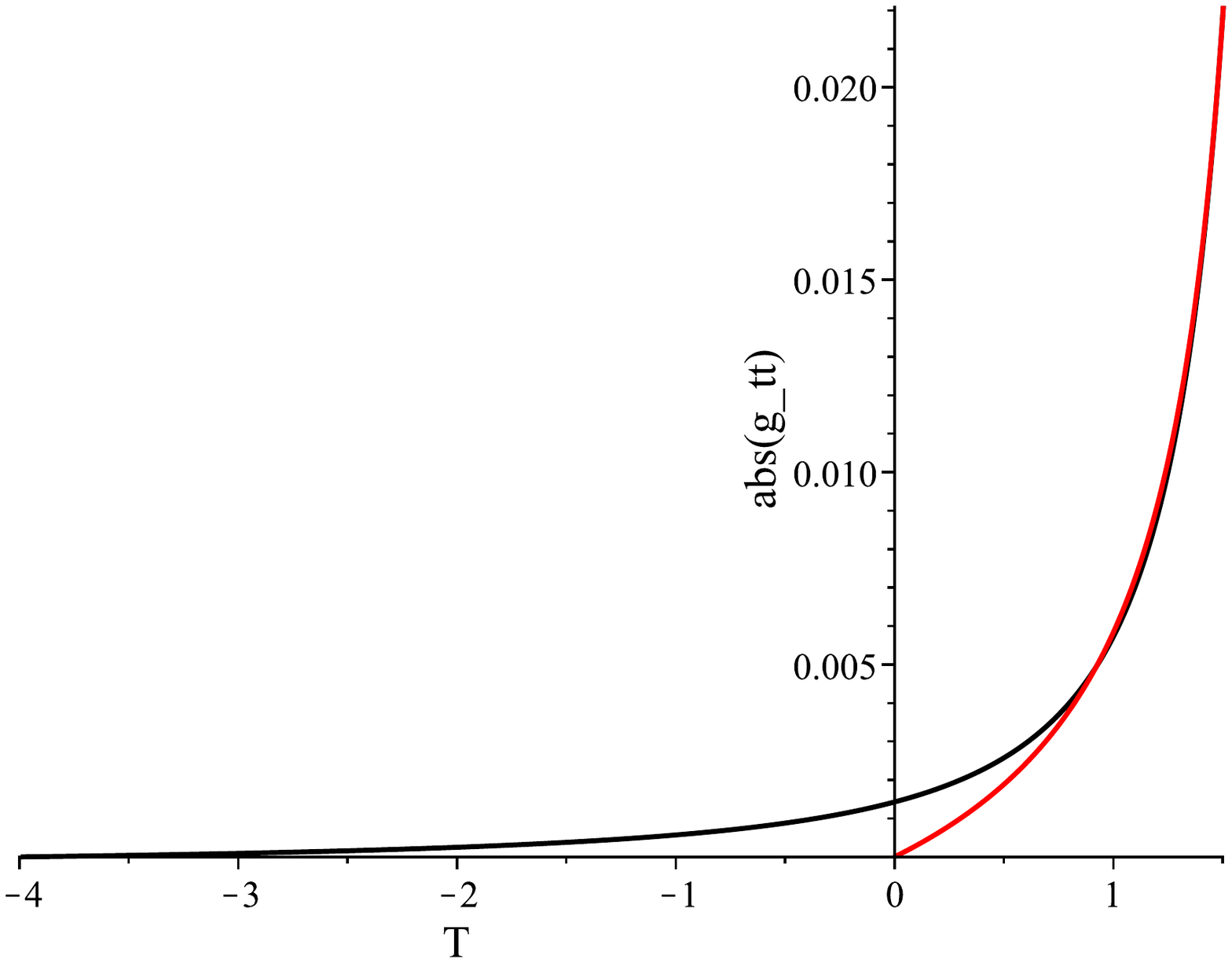}
\vspace{-5.0cm}\caption{{\linespread{0.4}{\footnotesize{$\theta\neq 0,\,\beta=0$. com\-mut\-a\-tive (red) vs non-com\-mut\-a\-tive (black) interior of a spherical black hole. Top: The triad component $\Ethree$. Middle: The triad combination $(\Etwo)^2/\Ethree$, corresponding to $g_{yy}$. Bottom: $N^{2}$, corresponding to $|g_{\tau\tau}|$. Notice from the top graph that the radius of the 2D subspaces, governed by the value of $\Ethree$, shrinks to zero beyond $\tau=0$ in the non-com\-mut\-a\-tive case, indicating a delay in the singularity. The parameters are: $M=1$, $\Lambda=-0.1$, $\gamma=0.274$, $\theta=-0.4$, and $\beta=0$.}}}}
\label{fig:thetasphds}
\end{figure}

\begin{figure}[h!t]
\centering
\vspace{0.0cm} \includegraphics[width=1.0\columnwidth, clip]{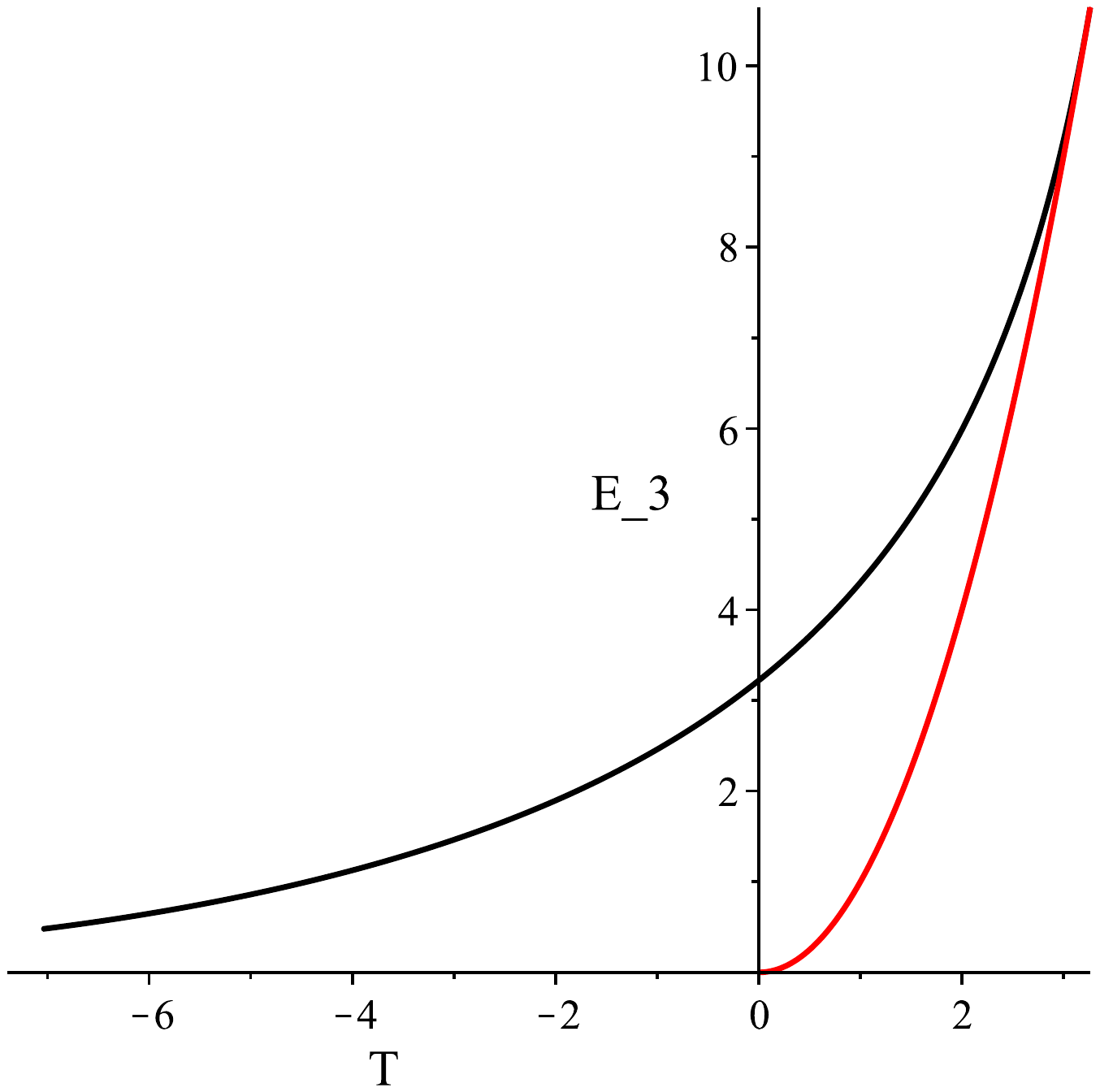}\\[-5.0cm]
\hspace{-0.3cm} \includegraphics[width=1.0\columnwidth, clip]{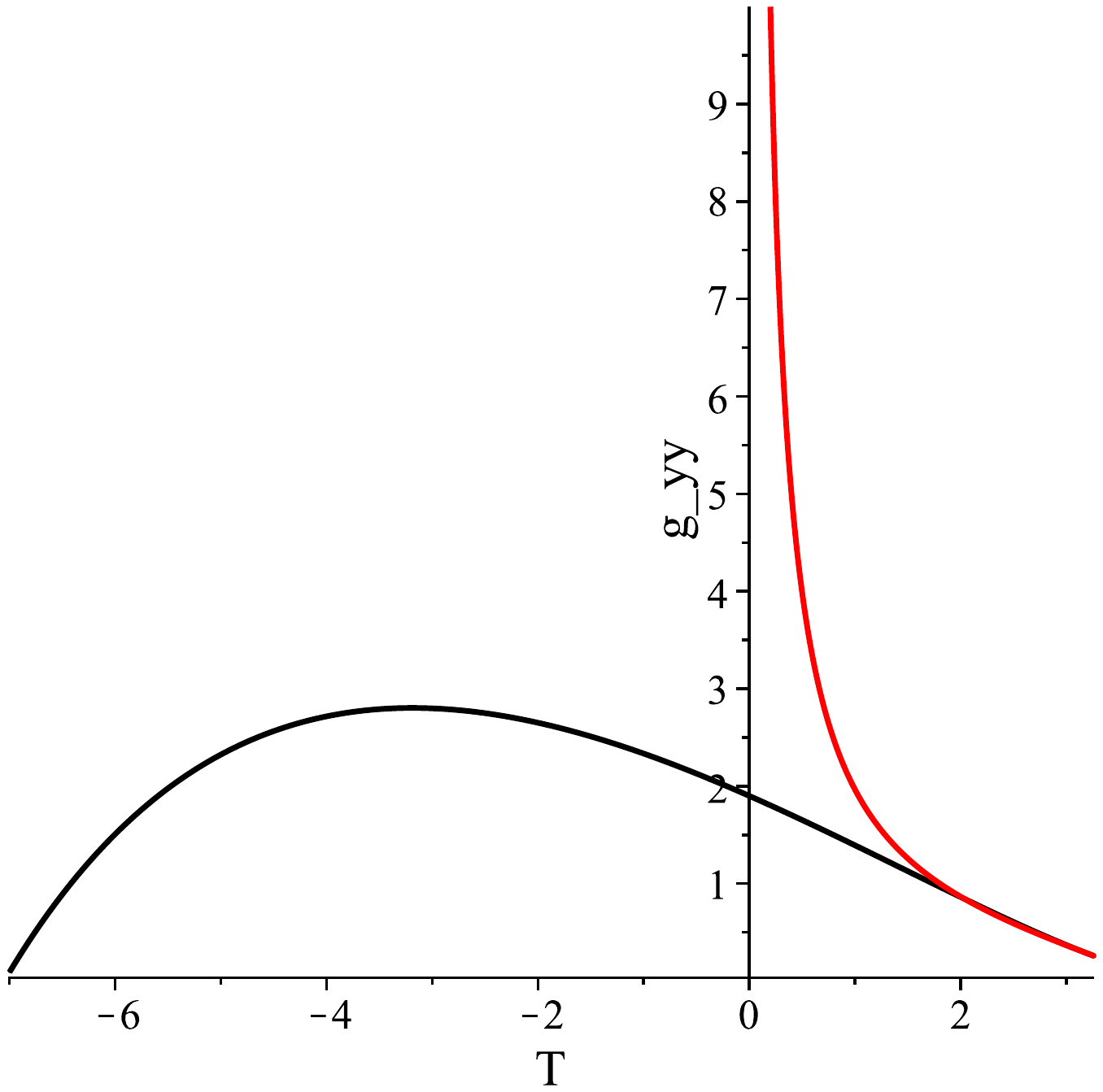}\\[-5.0cm]
\hspace{-0.3cm}\includegraphics[width=1.0\columnwidth, clip]{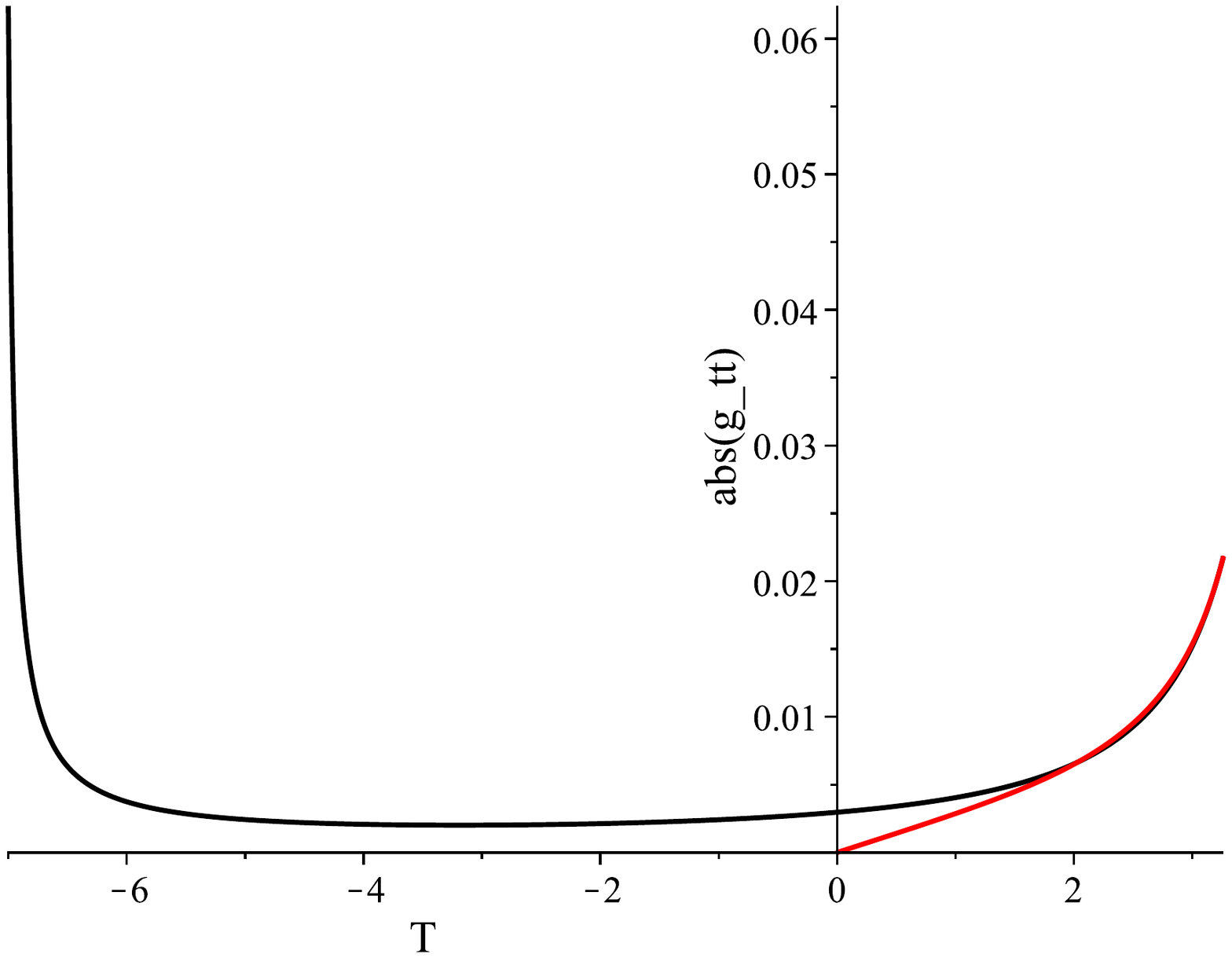}
\vspace{-5.0cm}\caption{{\linespread{0.4}{\footnotesize{$\theta\neq 0,\,\beta=0$. com\-mut\-a\-tive (red) vs non-com\-mut\-a\-tive (black) interior of a toroidal black hole. Top: The triad component $\Ethree$. Middle: The triad combination $(\Etwo)^2/\Ethree$, corresponding to $g_{yy}$. Bottom: $N^{2}$, corresponding to $|g_{\tau\tau}|$. Notice from the top graph that the radius of the 2D subspaces in this particular case does \emph{not} shrink to zero in the non-com\-mut\-a\-tive case, indicating the removal of the singularity, but the evolution stops due to $g_{yy}\rightarrow 0$ and $|g_{\tau\tau}| \rightarrow \infty$, indicating the presence of another horizon. The parameters are: $M=1$, $\Lambda=-0.1$, $\gamma=0.274$, $\theta=-0.08$, and $\beta=0$.}}}}
\label{fig:thetatords}
\end{figure}

\begin{figure}[h!t]
\centering
\vspace{0.0cm} \includegraphics[width=1.0\columnwidth, clip]{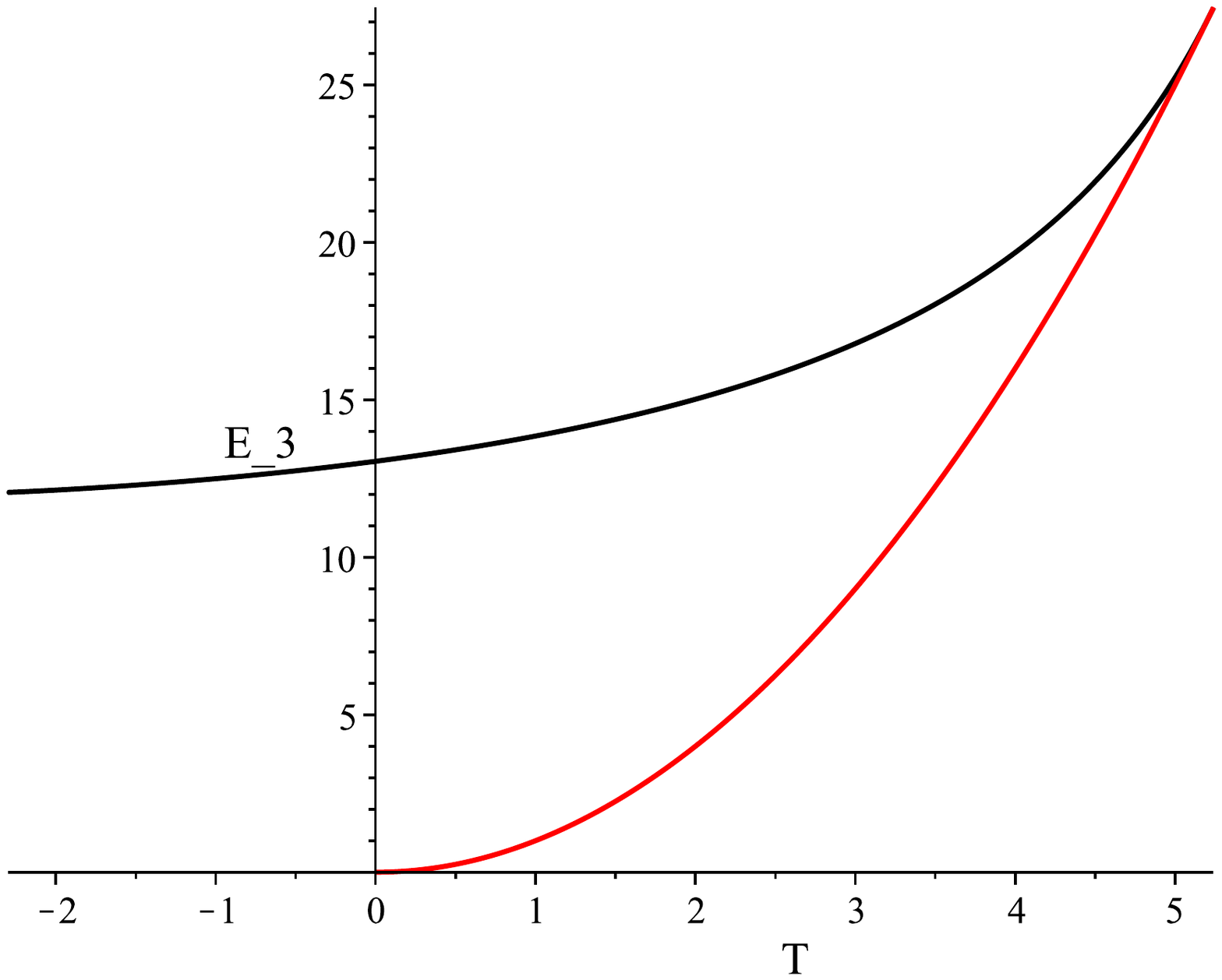}\\[-5.0cm]
\hspace{-0.3cm} \includegraphics[width=1.0\columnwidth, clip]{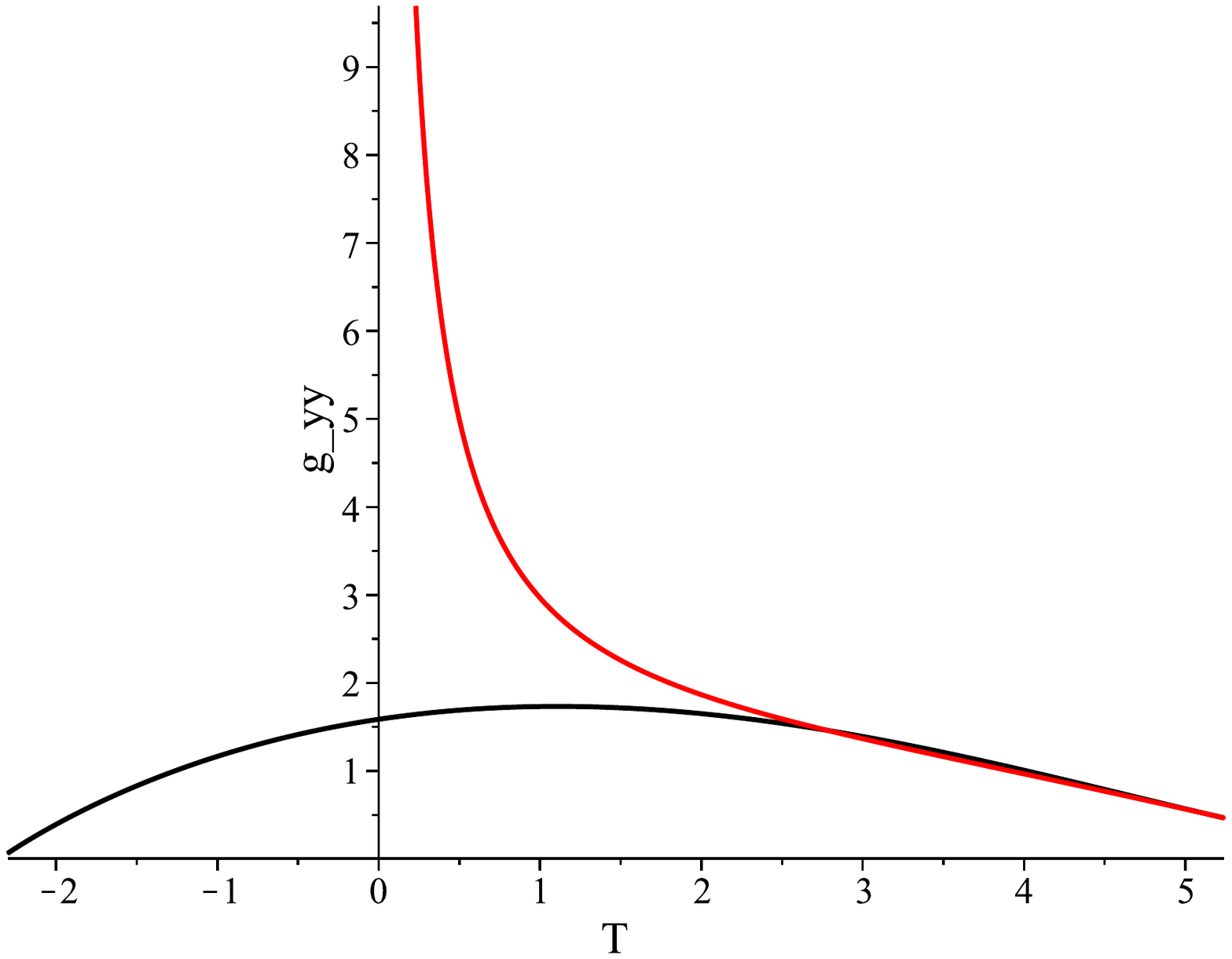}\\[-5.0cm]
\hspace{-0.3cm}\includegraphics[width=1.0\columnwidth, clip]{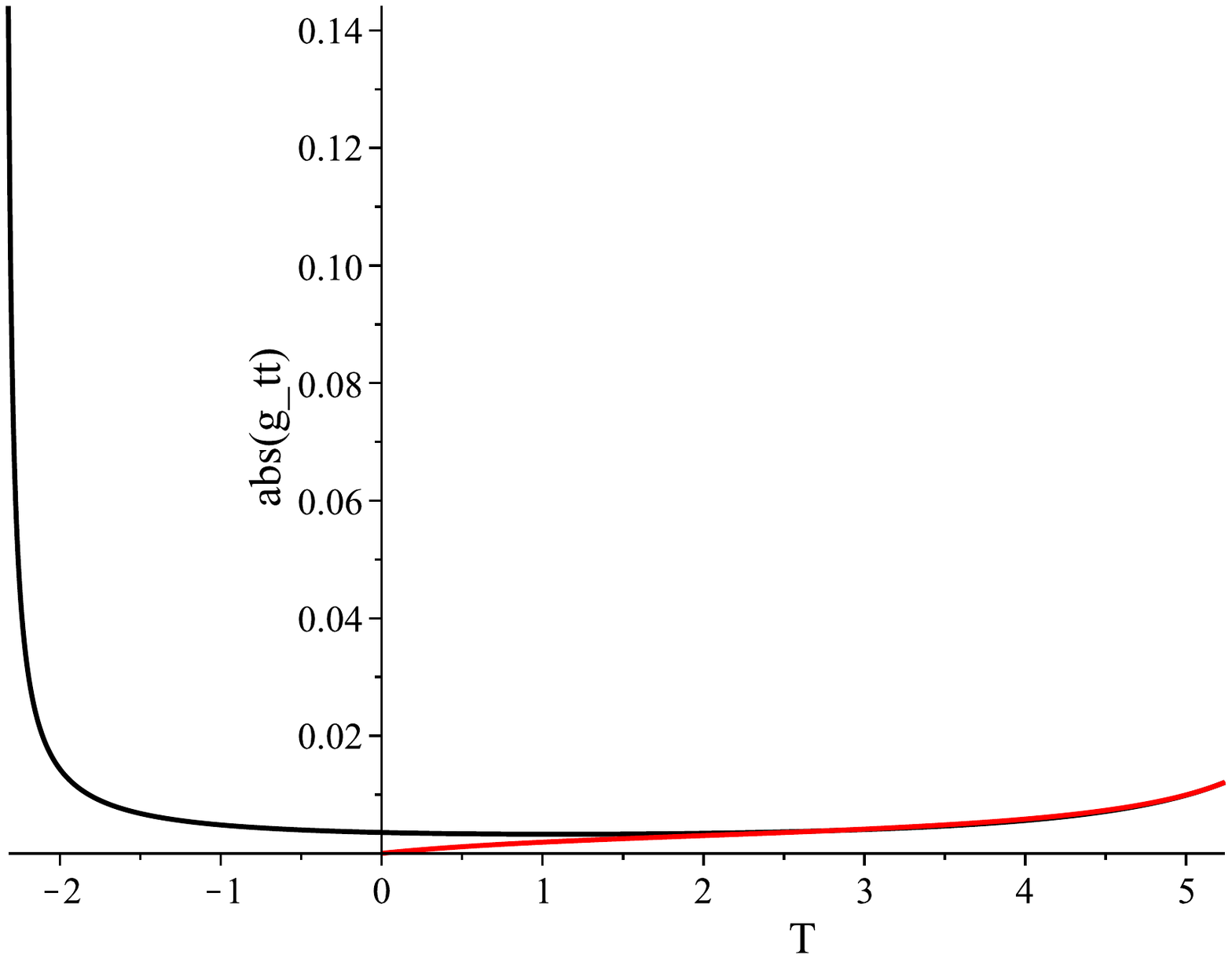}
\vspace{-5.0cm}\caption{\vspace{0.0cm}{\linespread{0.4}{\footnotesize{$\theta\neq 0,\,\beta=0$. com\-mut\-a\-tive (red) vs non-com\-mut\-a\-tive (black) interior of a higher-genus black hole. Top: The triad component $\Ethree$. Middle: The triad combination $(\Etwo)^2/\Ethree$, corresponding to $g_{yy}$. Bottom: $N^{2}$, corresponding to $|g_{\tau\tau}|$. Notice from the top graph that the radius of the 2D subspaces in this particular case does \emph{not} shrink to zero in the non-com\-mut\-a\-tive case, but the evolution stops due to $g_{yy}\rightarrow 0$ and $|g_{\tau\tau}| \rightarrow \infty$, indicating the presence of another horizon. The parameters are: $M=1$, $\Lambda=-0.1$, $\gamma=0.274$, $\theta=-0.06$, and $\beta=0$.\vspace{0.0cm}}}}}
\label{fig:thetahypds}
\end{figure}

\enlargethispage{0.7cm}The figures show a few, but not all, possible scenarios and below in table 1 we summarize all cases. In none of the solutions can the results be evolved indefinitely. In some cases shown there is a true singularity present, with $\Ethree$ shrinking to zero, whereas in others it is a (curvature) finite solution ($\Ethree$ non-zero) but with a new horizon appearing (see figure captions for details)\footnote{The condition for an event horizon may seem peculiar here but note that in the $T$-domain the metric condition
 $g_{yy}\rightarrow 0_{+}$ is an equivalent statement to $g_{tt} \rightarrow 0_{-}$ ($t$ being \emph{exterior} time).}. In general it is found that if $\beta=0$, and the non-com\-mut\-a\-tiv\-ity parameter, $\theta$ is fairly large, one may eliminate the singularity, although in some cases an inner horizon results, and we are unable to probe beyond that horizon to glean if there is singular structure hiding behind it. For small enough values of $\theta$ the singularity is always present. 

\subsubsection{Non-com\-mut\-a\-tive triad only}
For the case where the connection remains self-com\-mut\-a\-tive but the triad becomes non-com\-mut\-a\-tive we present the sets of scenarios in figures \ref{fig:betasphds}-\ref{fig:betahypds}, along with a full summary in table 1.

\begin{figure}[h!t]
\centering
\vspace{0.0cm} \includegraphics[width=1.0\columnwidth, clip]{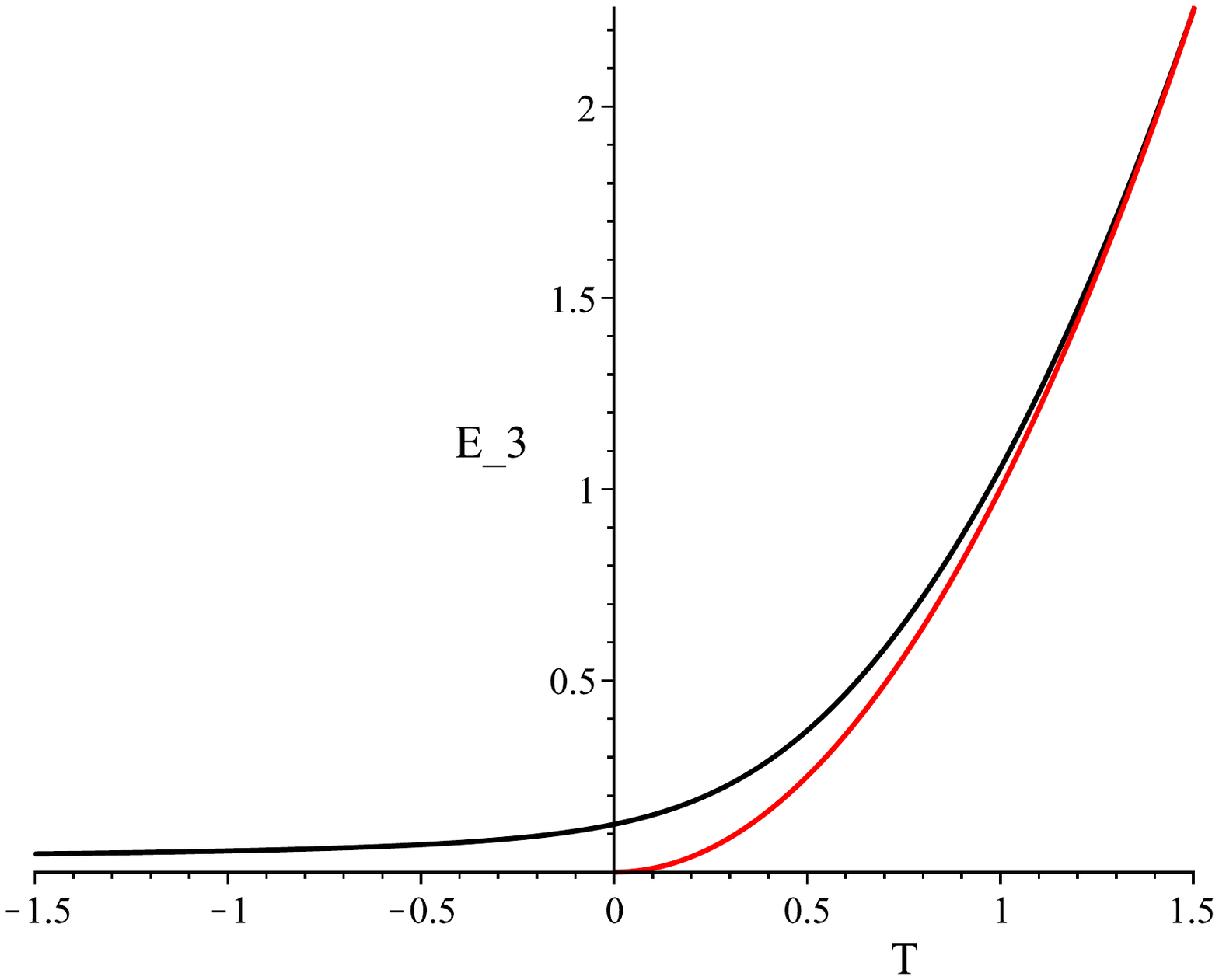}\\[-5.0cm]
\hspace{-0.3cm} \includegraphics[width=1.0\columnwidth, clip]{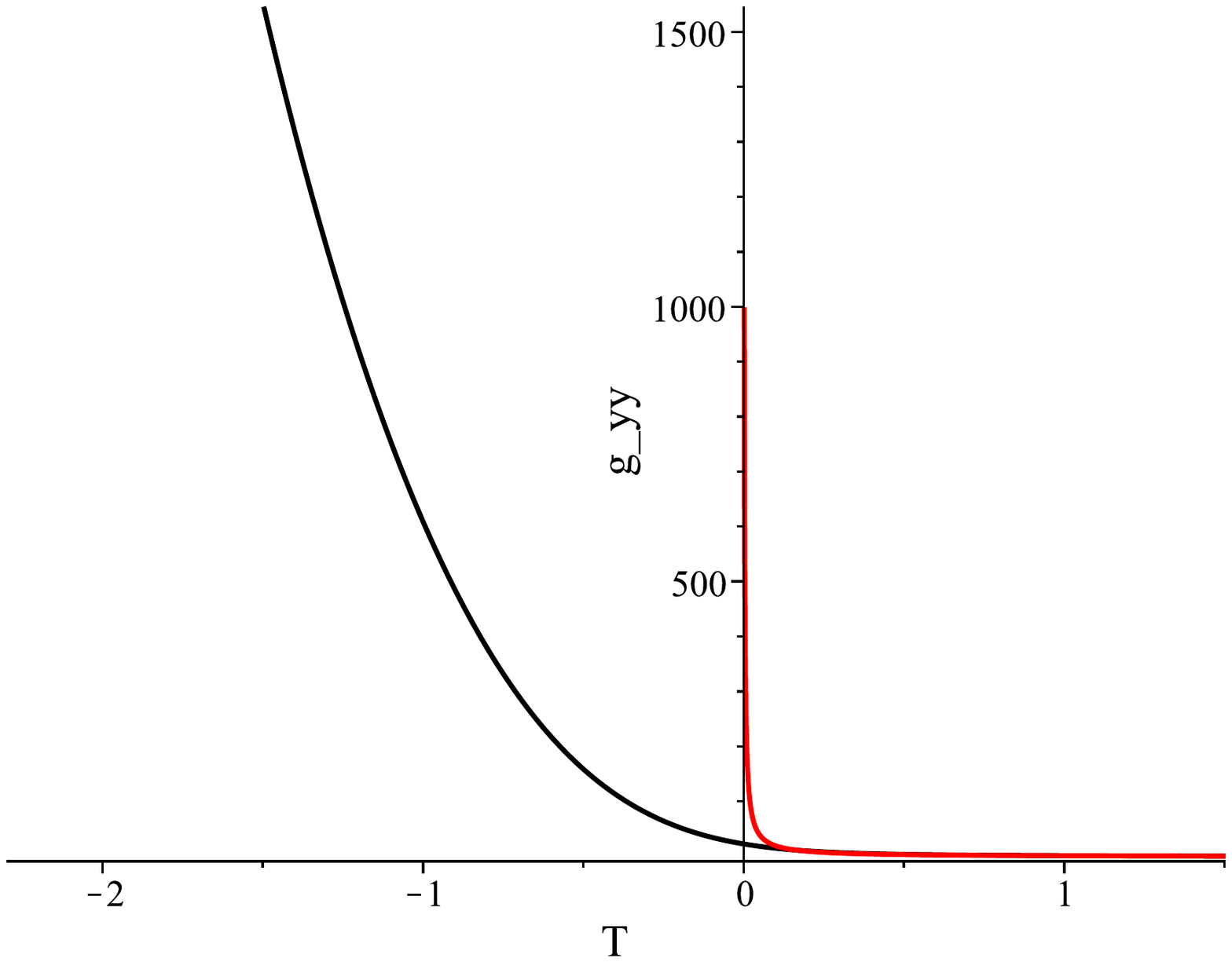}\\[-5.0cm]
\hspace{-0.3cm}\includegraphics[width=1.0\columnwidth, clip]{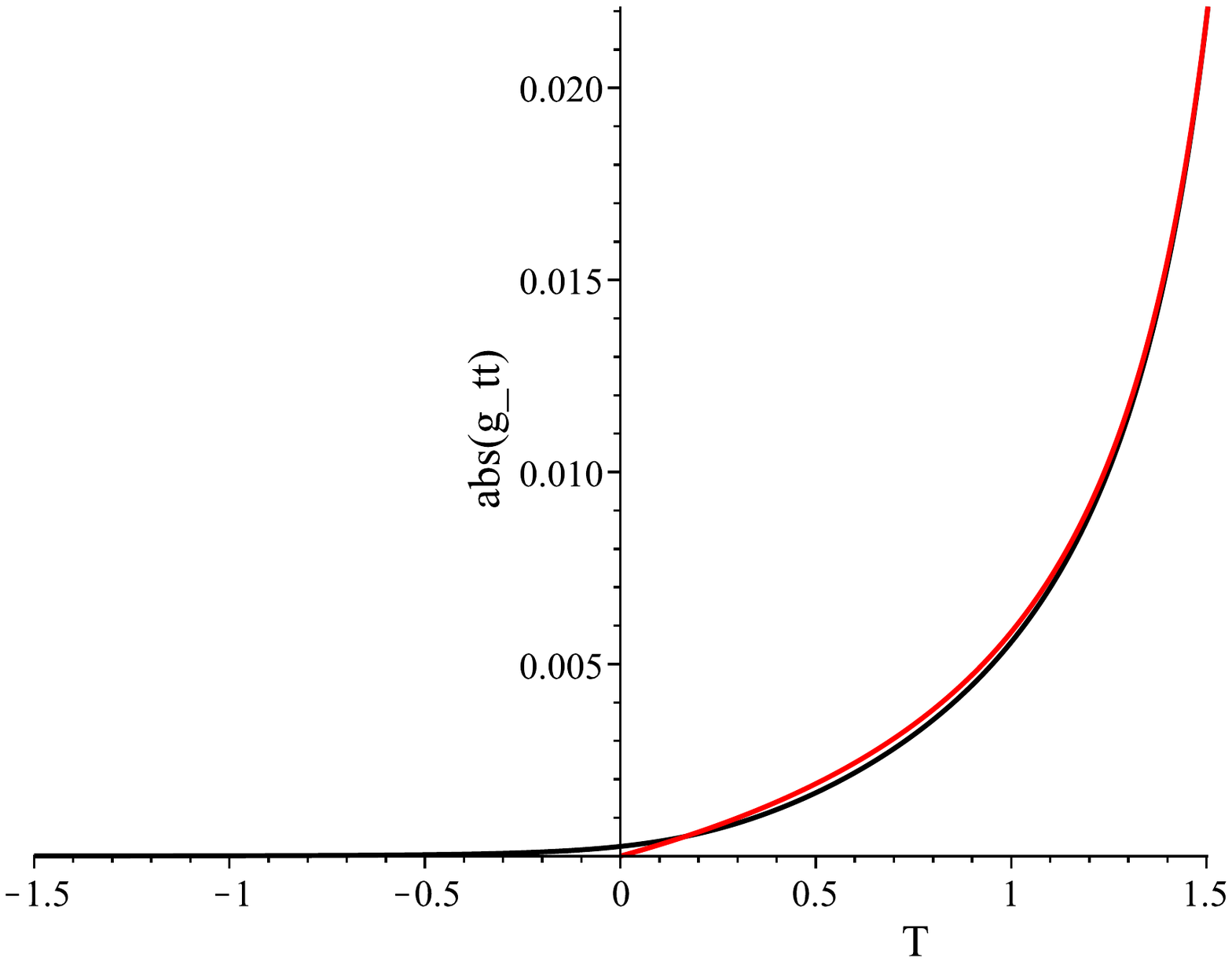}
\vspace{-5.0cm}\caption{{\linespread{0.4}{\footnotesize{$\theta=0,\,\beta\neq 0$. com\-mut\-a\-tive (red) vs non-com\-mut\-a\-tive (black) interior of a spherical black hole. Top: The triad component $\Ethree$. Middle: The triad combination $(\Etwo)^2/\Ethree$, corresponding to $g_{yy}$. Bottom: $N^{2}$, corresponding to $|g_{\tau\tau}|$. Notice from the top graph that the radius of the 2D subspaces, governed by the value of $\Ethree$, does \emph{not} shrink to zero in the non-com\-mut\-a\-tive case, indicating removal of the singularity. The parameters are: $M=1$, $\Lambda=-0.1$, $\gamma=0.274$, $\theta=0$, and $\beta=-0.1$. There is no new horizon in this case.}}}}
\label{fig:betasphds}
\end{figure}

\begin{figure}[h!t]
\centering
\vspace{0.0cm} \includegraphics[width=1.0\columnwidth, clip]{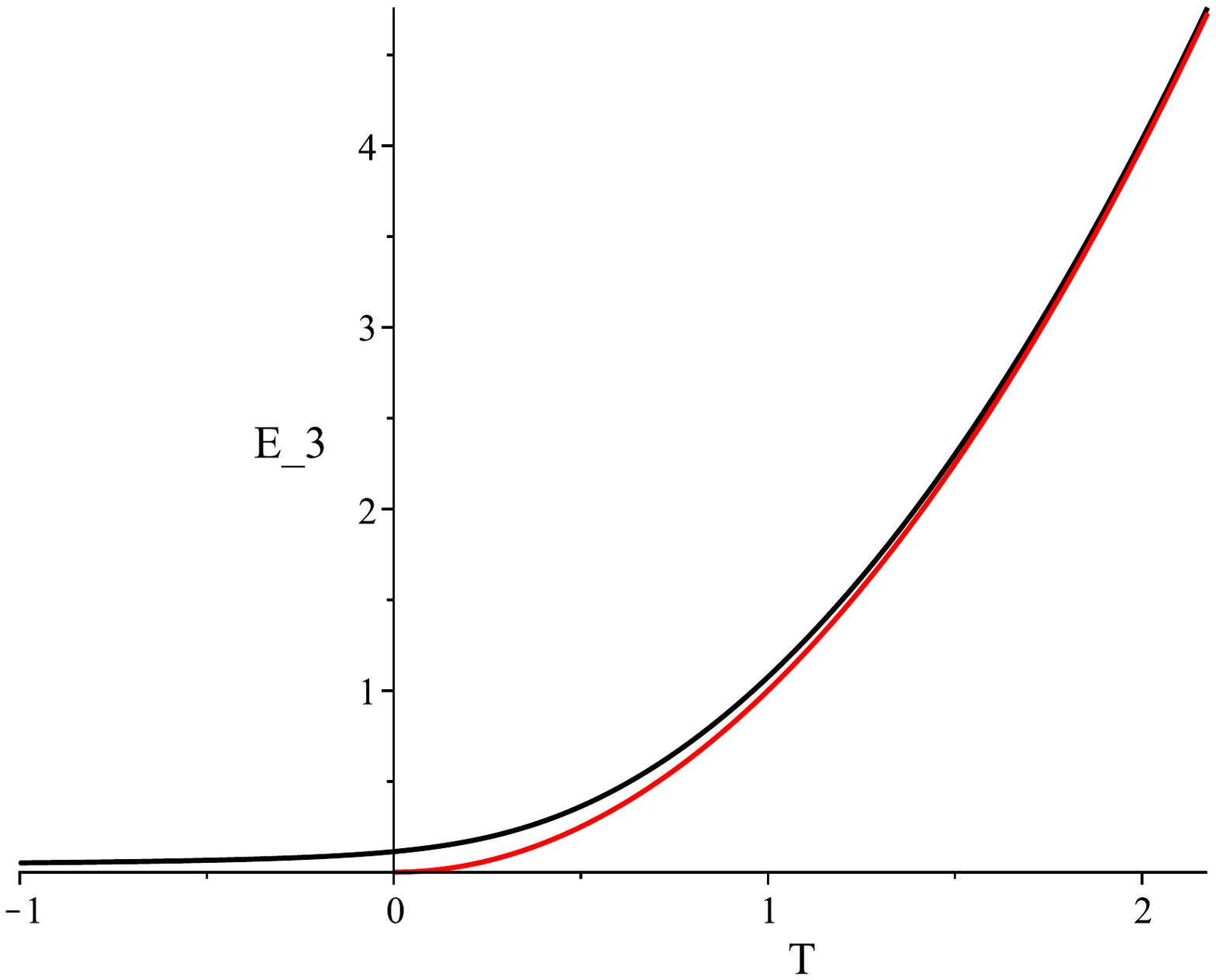}\\[-5.0cm]
\hspace{-0.3cm} \includegraphics[width=1.0\columnwidth, clip]{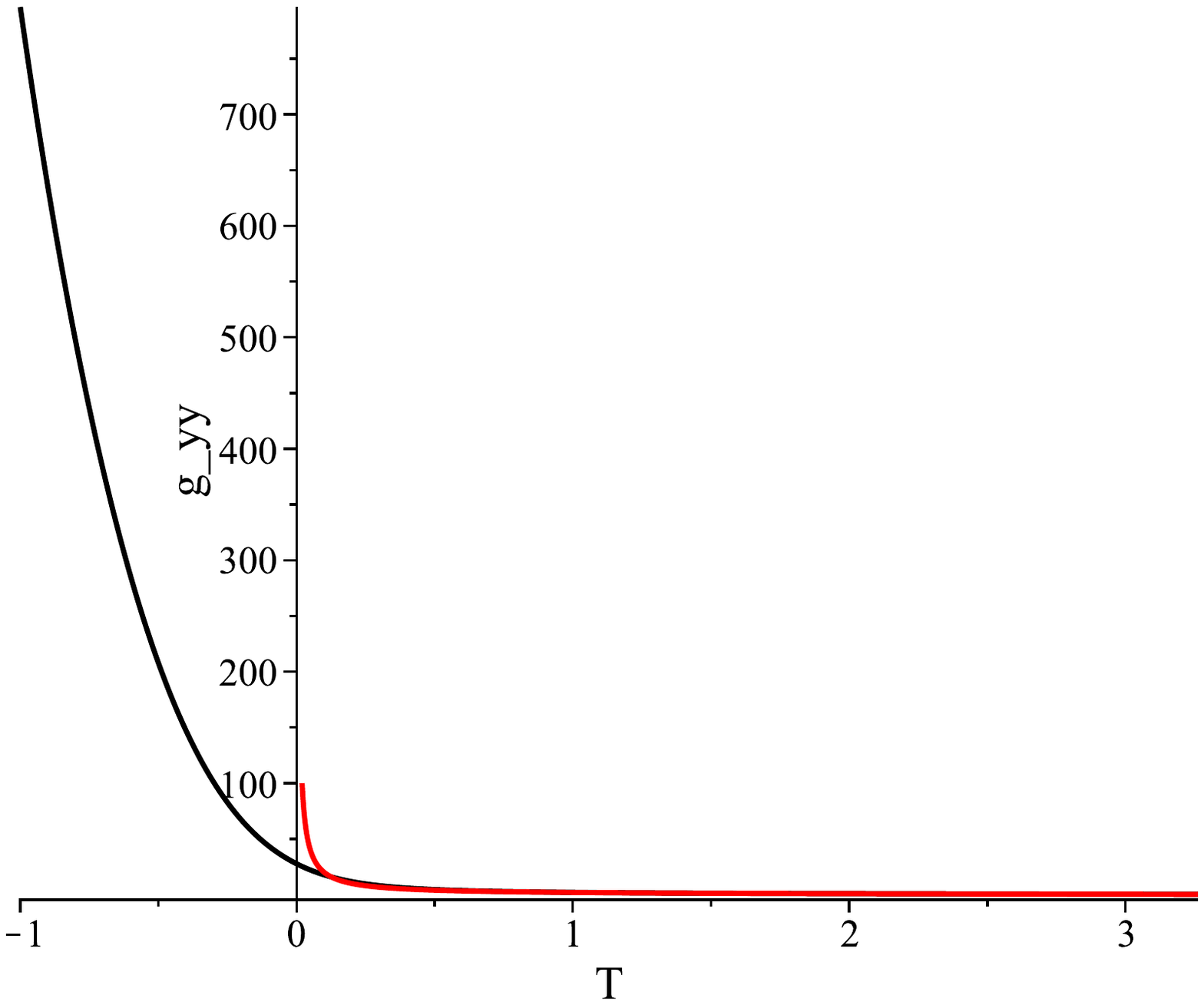}\\[-5.0cm]
\hspace{-0.3cm}\includegraphics[width=1.0\columnwidth, clip]{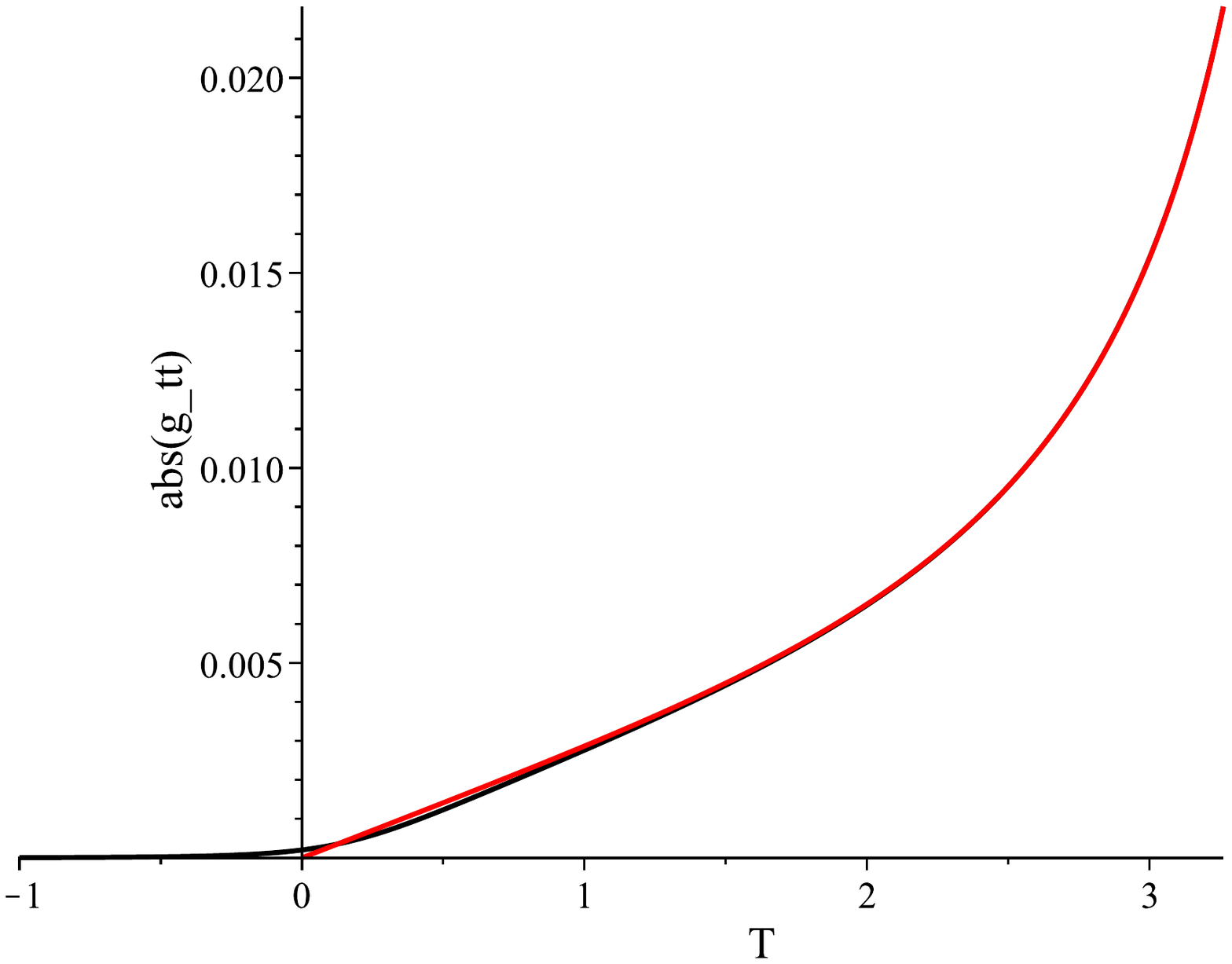}
\vspace{-5.0cm}\caption{{\linespread{0.4}{\footnotesize{$\theta=0,\,\beta\neq 0$. com\-mut\-a\-tive (red) vs non-com\-mut\-a\-tive (black) interior of a toroidal black hole. Top: The triad component $\Ethree$. Middle: The triad combination $(\Etwo)^2/\Ethree$, corresponding to $g_{yy}$. Bottom: $N^{2}$, corresponding to $|g_{\tau\tau}|$. Notice from the top graph that the radius of the 2D subspaces does \emph{not} shrink to zero in the non-com\-mut\-a\-tive case, indicating removal of the singularity. The parameters are: $M=1$, $\Lambda=-0.1$, $\gamma=0.274$, $\theta=0$, and $\beta=-0.1$. There is no new horizon in this case.}}}}
\label{fig:betatords}
\end{figure}

\begin{figure}[h!t]
\centering
\vspace{0.0cm} \includegraphics[width=1.0\columnwidth, clip]{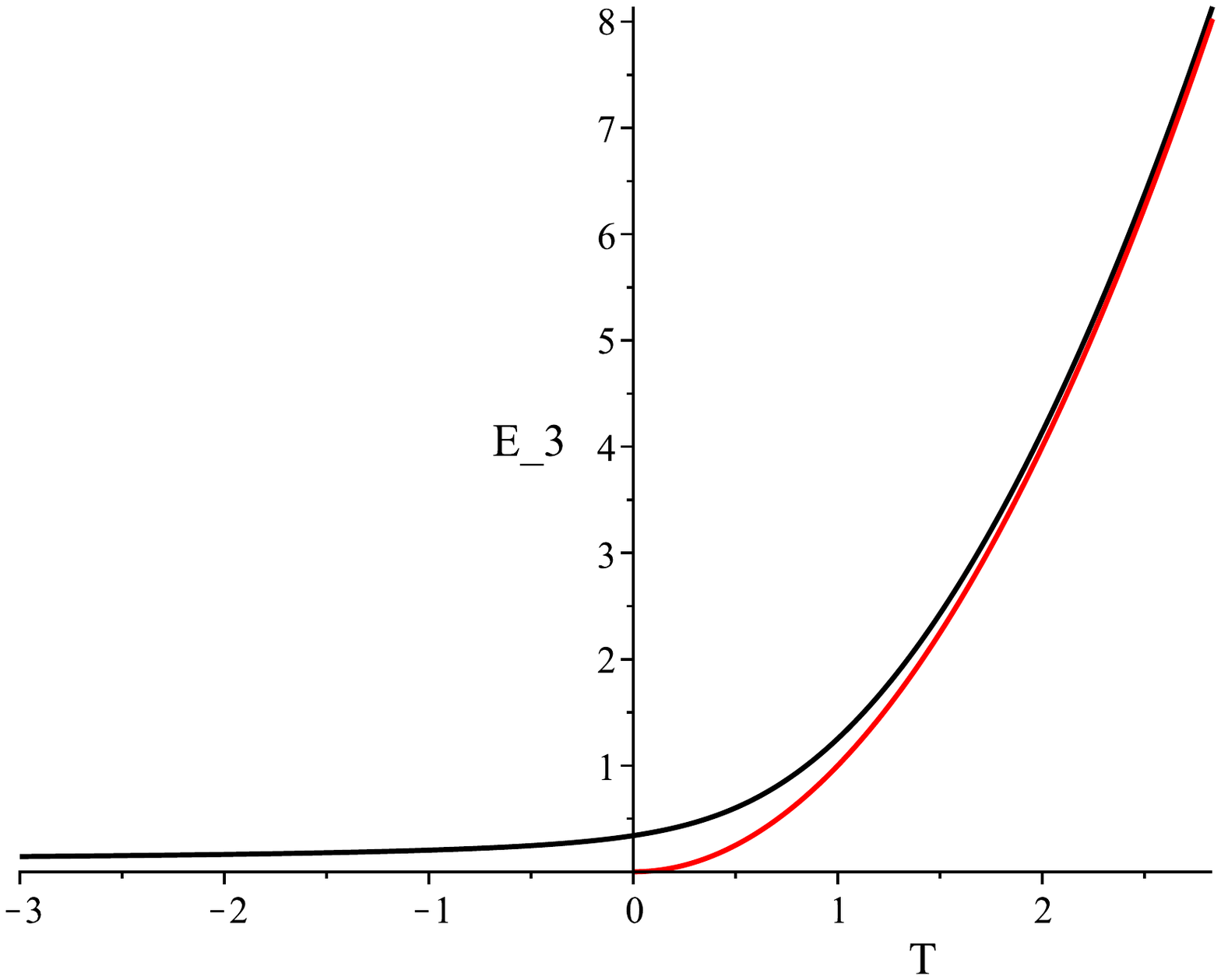}\\[-5.0cm]
\hspace{-0.3cm} \includegraphics[width=1.0\columnwidth, clip]{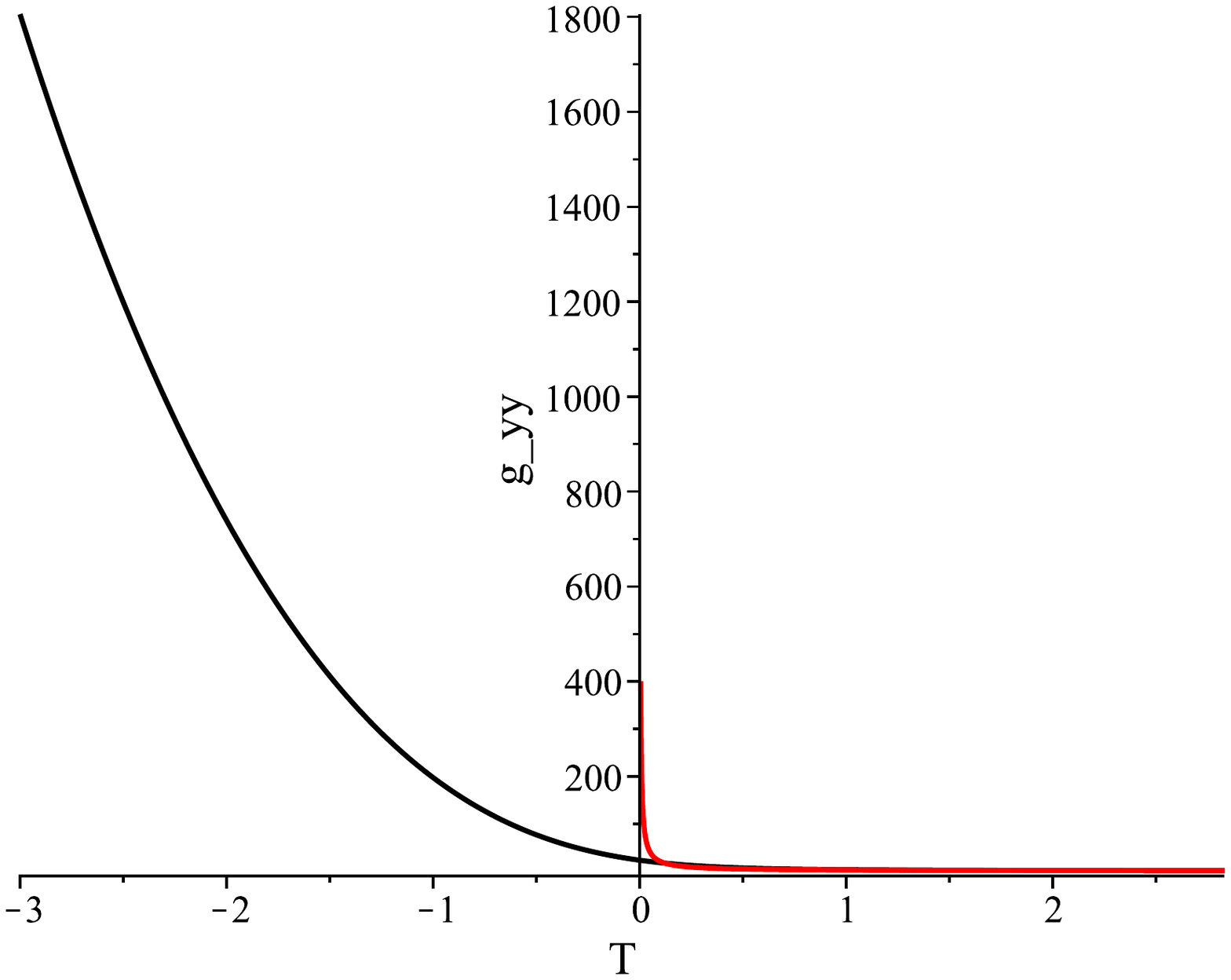}\\[-5.0cm]
\hspace{-0.3cm}\includegraphics[width=1.0\columnwidth, clip]{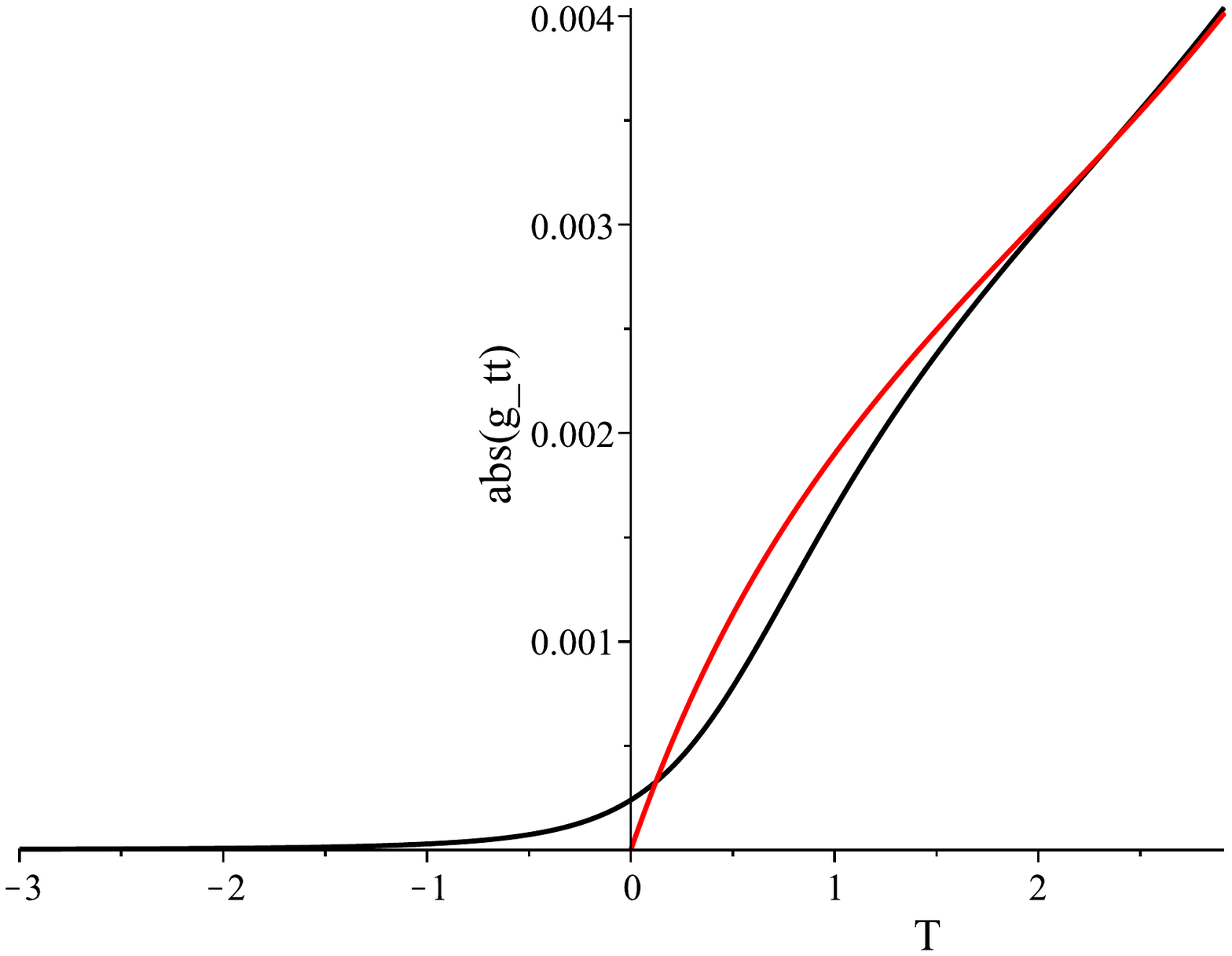}
\vspace{-5.0cm}\caption{{\linespread{0.4}{\footnotesize{$\theta=0,\,\beta\neq 0$. com\-mut\-a\-tive (red) vs non-com\-mut\-a\-tive (black) interior of a higher-genus black hole. Top: The triad component $\Ethree$. Middle: The triad combination $(\Etwo)^2/\Ethree$, corresponding to $g_{yy}$. Bottom: $N^{2}$, corresponding to $|g_{\tau\tau}|$. Notice from the top graph that the radius of the 2D subspaces does \emph{not} shrink to zero in the non-com\-mut\-a\-tive case, indicating removal of the singularity. The parameters are: $M=1$, $\Lambda=-0.1$, $\gamma=0.274$, $\theta=0$, and $\beta=-0.5$. There is no new horizon in this case.}}}}
\label{fig:betahypds}
\end{figure}

It turns out that for both large and small non-zero values of $\beta$ the quantity $\Ethree$ asymptotes to a non-zero constant. The size of the 2D subspaces, governed by the value of $\Ethree$, depends on the value of $\beta$, with larger $\beta$ values yielding larger volumes. The situation here is somewhat reminiscent of what occurs in effective loop quantum gravity when holonomy corrections are introduced, the main difference being that in the loop quantum gravity scenario the volume of the subspaces oscillates in a damped manner, asymptotically approaching a constant for large negative $\tau$ \cite{ref:ourlambdabh}. Although some appear small in the plots, no non-com\-mut\-a\-tive metric component goes to zero in figures \ref{fig:betasphds}-\ref{fig:betahypds} and this remains true as long as $\beta \neq 0$.

\subsubsection{Non-com\-mut\-a\-tive connection and triad}
Finally, we consider here the scenarios where all quantities possess a non-trivial Poisson bracket. Some representative results are shown in figures \ref{fig:thetabetasphds}-\ref{fig:thetabetahypds}. (See figure captions for details.)

\begin{figure}[h!t]
\centering
\vspace{0.0cm} \includegraphics[width=1.0\columnwidth, clip]{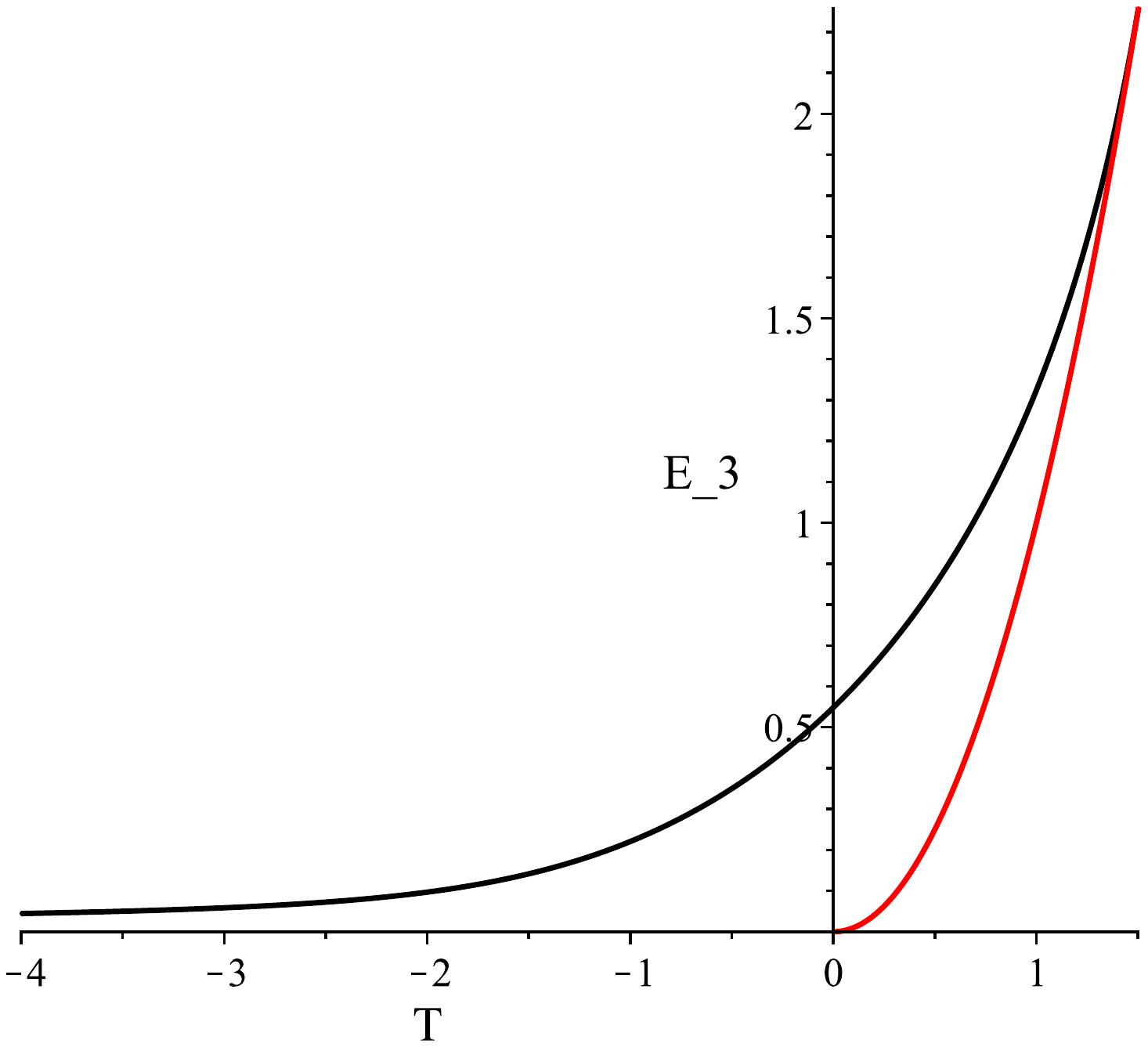}\\[-5.0cm]
\hspace{-0.3cm} \includegraphics[width=1.0\columnwidth, clip]{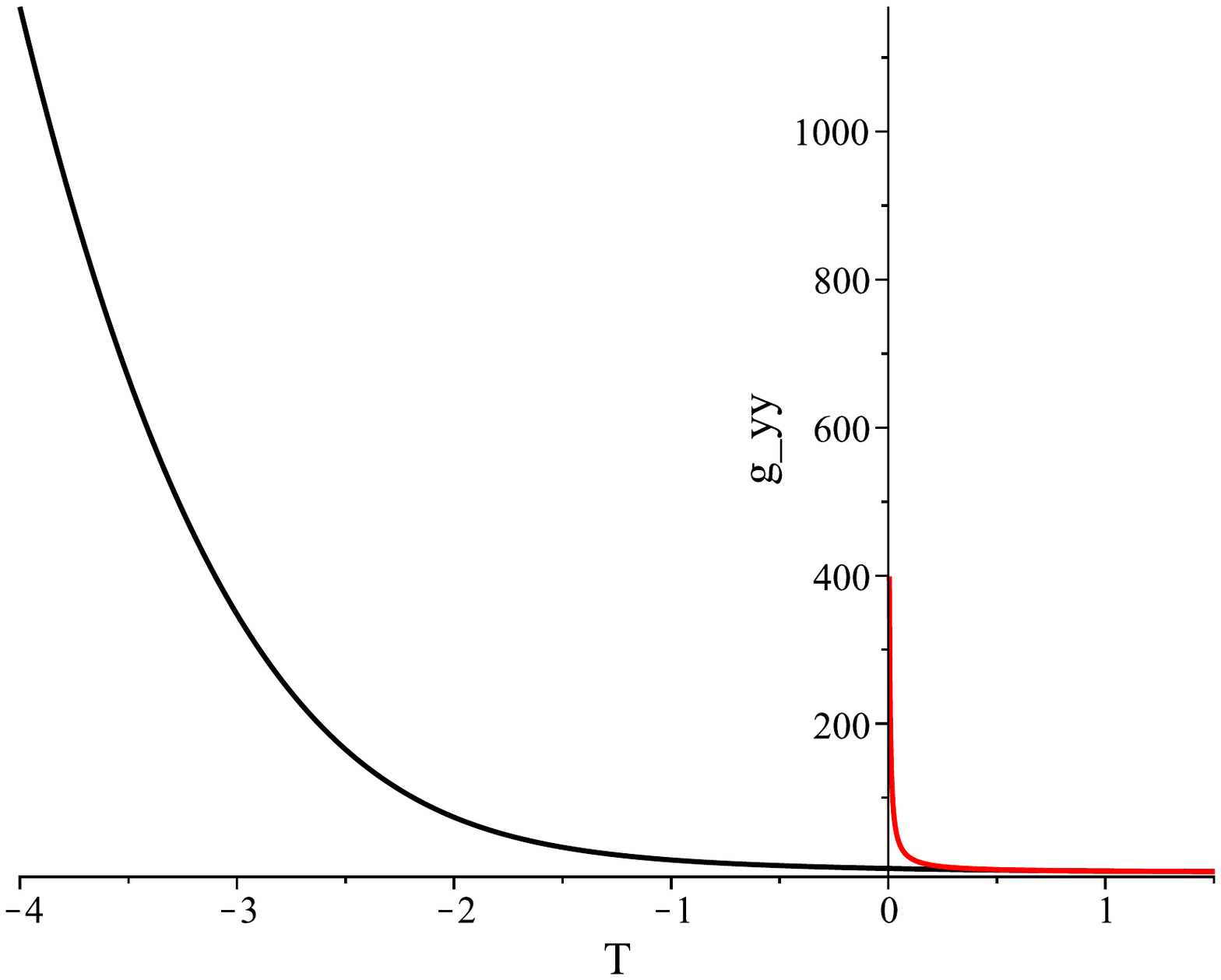}\\[-5.0cm]
\hspace{-0.3cm}\includegraphics[width=1.0\columnwidth, clip]{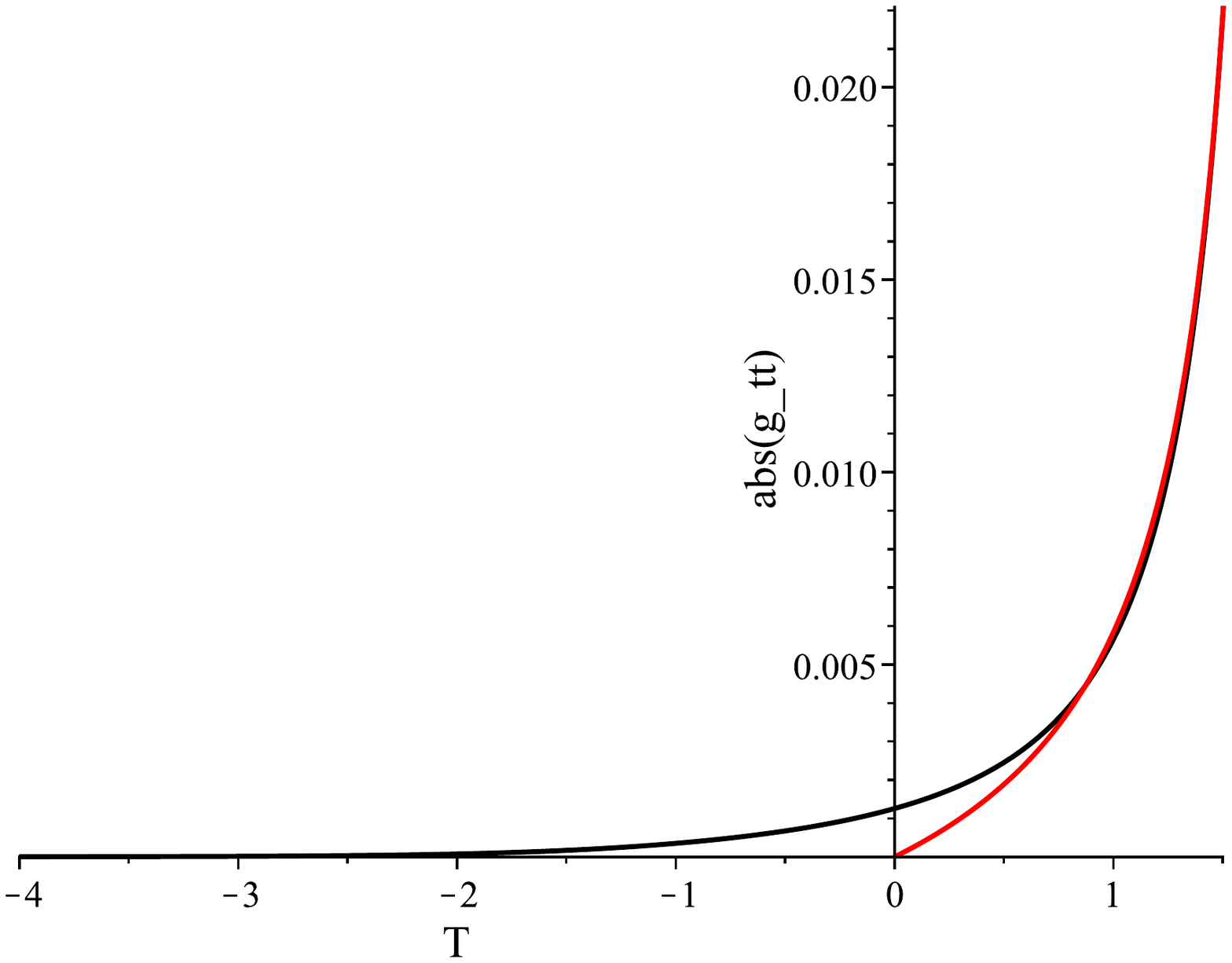}
\vspace{-5.0cm}\caption{{\linespread{0.4}{\footnotesize{$\theta\neq 0,\,\beta\neq 0$. com\-mut\-a\-tive (red) vs non-com\-mut\-a\-tive (black) interior of a spherical black hole. Top: The triad component $\Ethree$. Middle: The triad combination $(\Etwo)^2/\Ethree$, corresponding to $g_{yy}$. Bottom: $N^{2}$, corresponding to $|g_{\tau\tau}|$. Notice from the top graph that the radius of the 2D subspaces, governed by the value of $\Ethree$, does \emph{not} shrink to zero in the non-com\-mut\-a\-tive case, indicating removal of the singularity. The parameters are: $M=1$, $\Lambda=-0.1$, $\gamma=0.274$, $\theta=-0.3$, and $\beta=-0.1$. There is no new horizon in this case.}}}}
\label{fig:thetabetasphds}
\end{figure}

\begin{figure}[h!t]
\centering
\vspace{0.0cm} \includegraphics[width=1.0\columnwidth, clip]{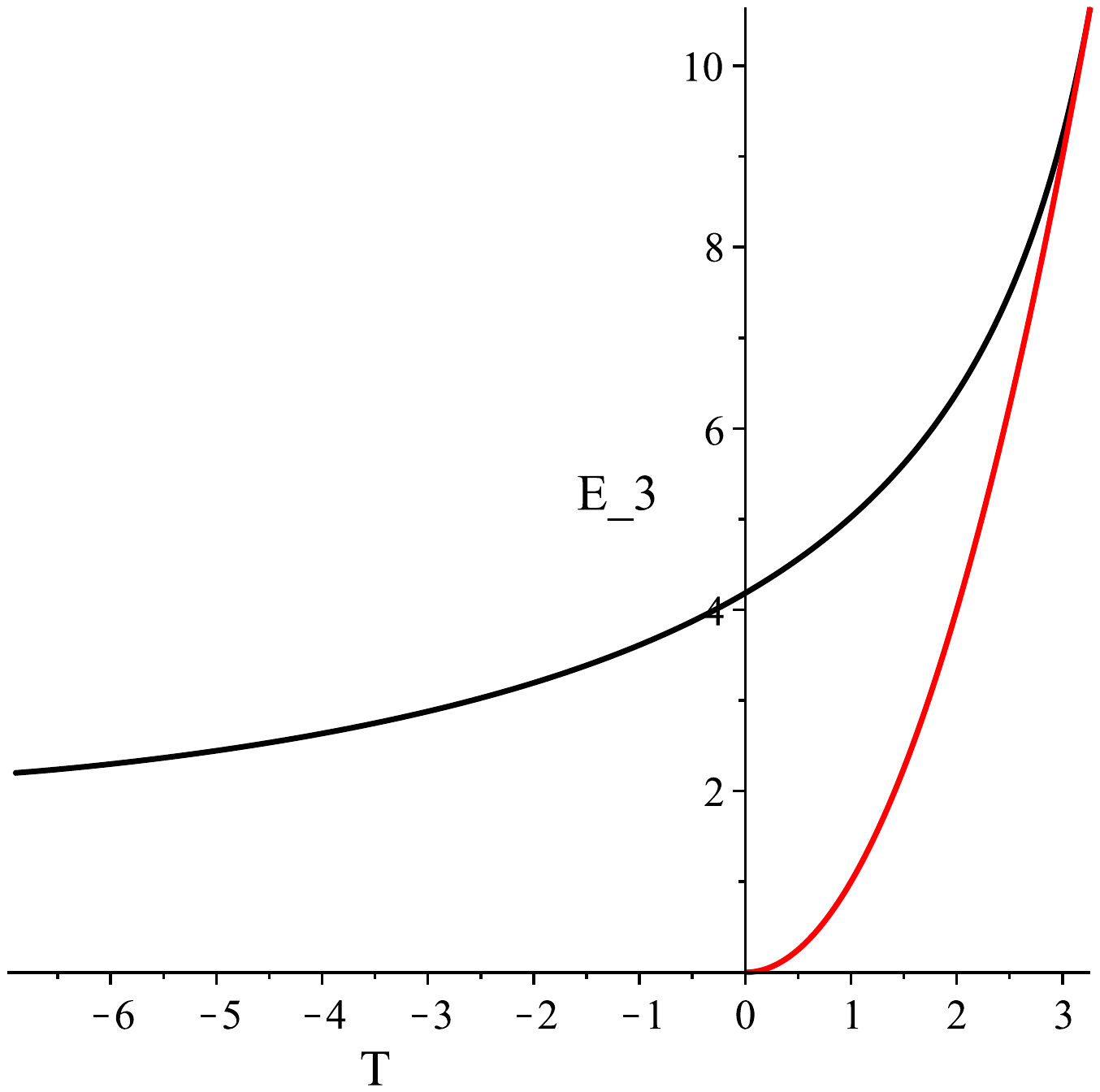}\\[-5.0cm]
\hspace{-0.3cm} \includegraphics[width=1.0\columnwidth, clip]{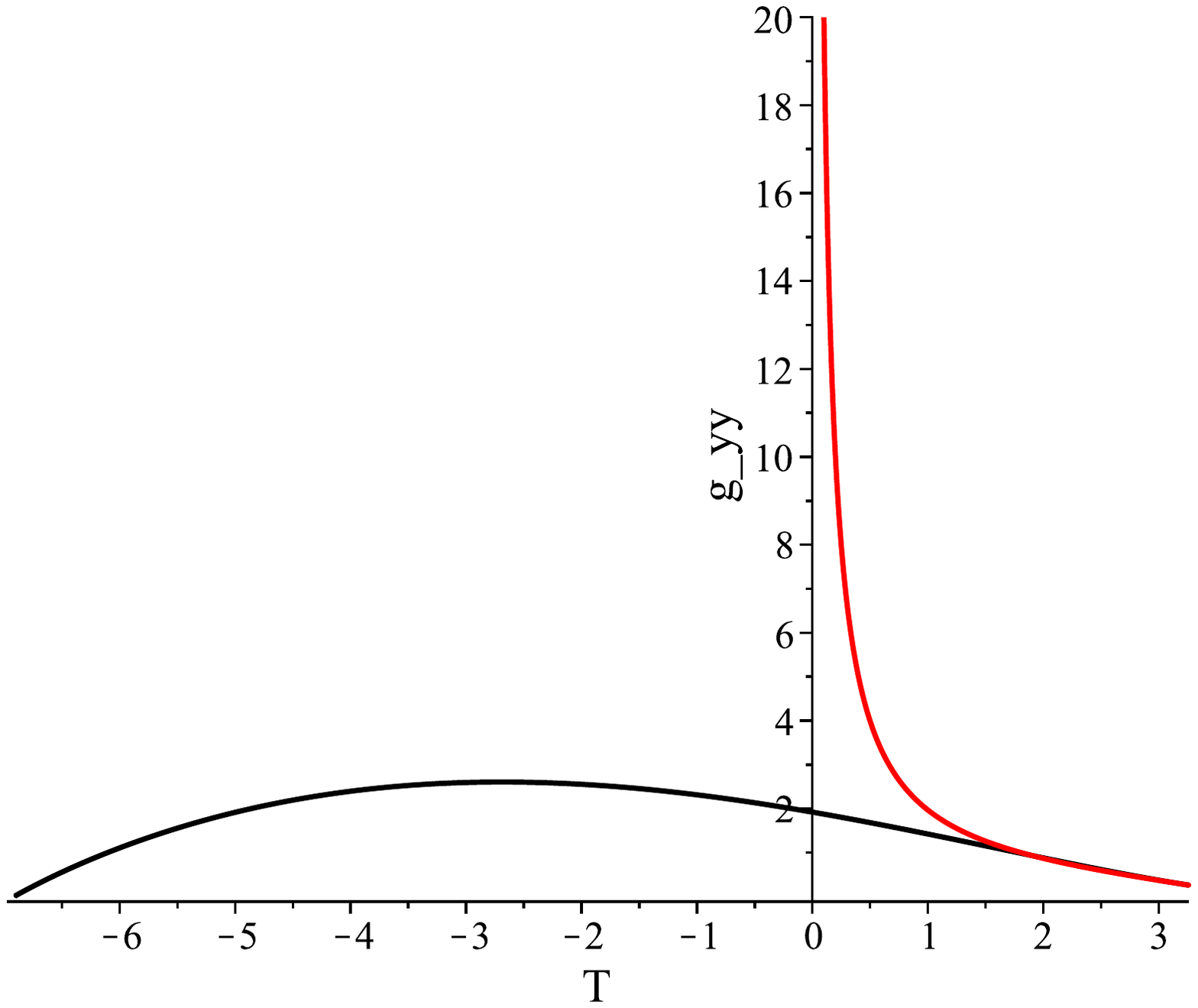}\\[-5.0cm]
\hspace{-0.3cm}\includegraphics[width=1.0\columnwidth, clip]{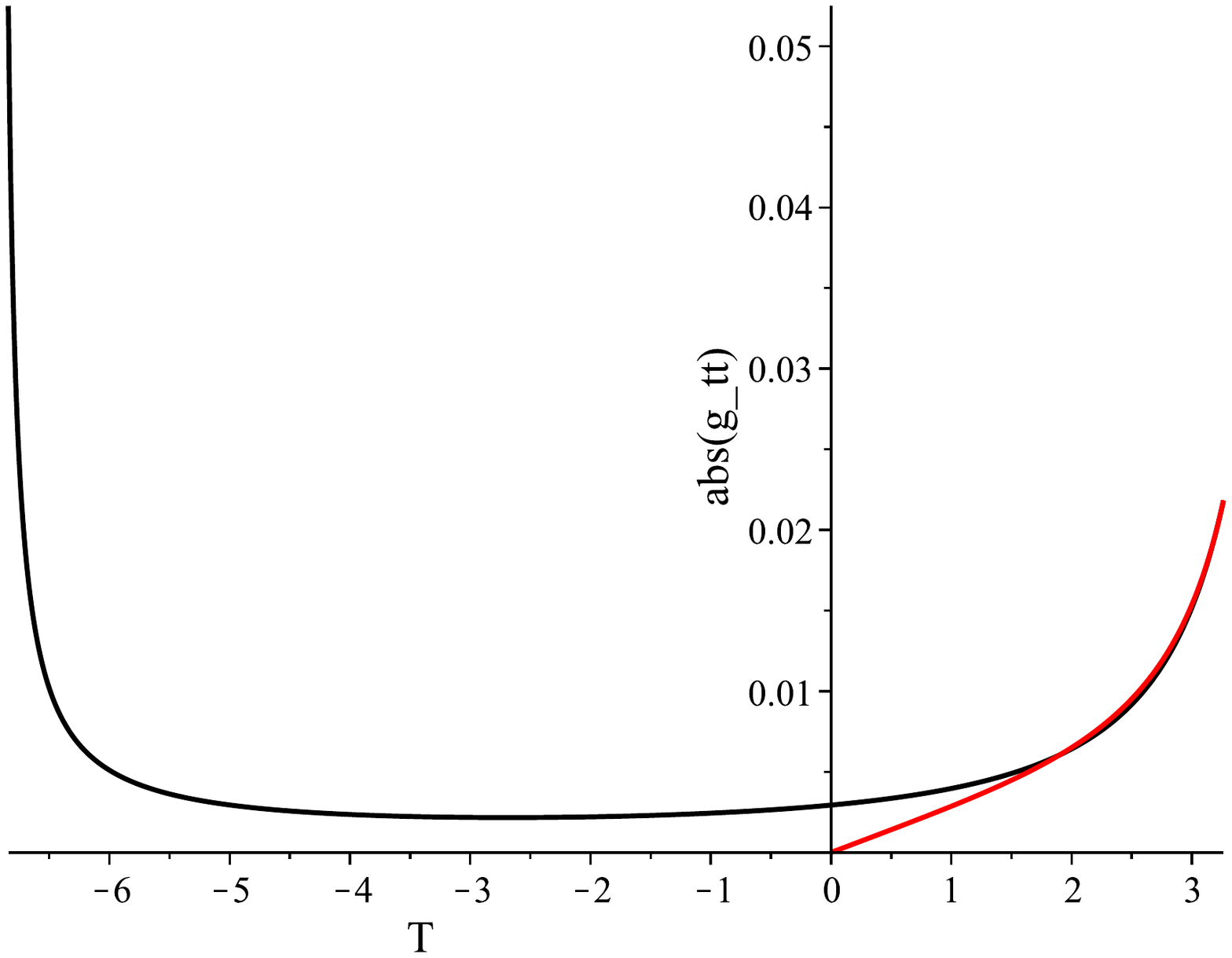}
\vspace{-5.0cm}\caption{{\linespread{0.4}{\footnotesize{$\theta\neq 0,\,\beta\neq 0$. com\-mut\-a\-tive (red) vs non-com\-mut\-a\-tive (black) interior of a toroidal black hole. Top: The triad component $\Ethree$. Middle: The triad combination $(\Etwo)^2/\Ethree$, corresponding to $g_{yy}$. Bottom: $N^{2}$, corresponding to $|g_{\tau\tau}|$. Notice from the top graph that the radius of the 2D subspaces in this particular case does \emph{not} shrink to zero in the non-com\-mut\-a\-tive case, indicating the removal of the singularity, but the evolution stops due to $g_{yy}\rightarrow 0$ and $|g_{\tau\tau}| \rightarrow \infty$, indicating the presence of another horizon. The parameters are: $M=1$, $\Lambda=-0.1$, $\gamma=0.274$, $\theta=-0.1$, and $\beta=-0.05$.}}}}
\label{fig:thetabetatords}
\end{figure}

\begin{figure}[h!t]
\centering
\vspace{0.0cm} \includegraphics[width=1.0\columnwidth, clip]{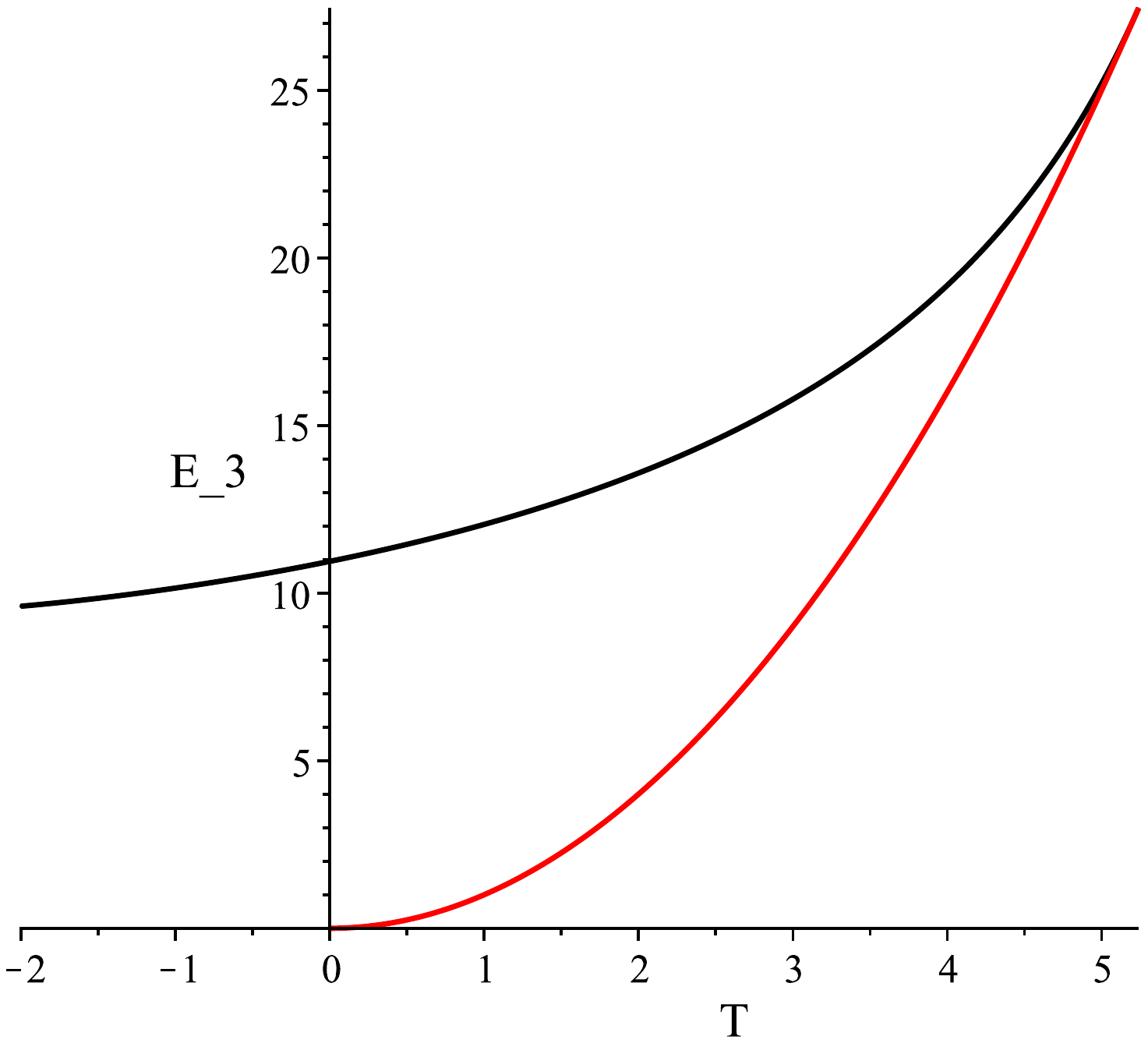}\\[-5.0cm]
\hspace{-0.3cm} \includegraphics[width=1.0\columnwidth, clip]{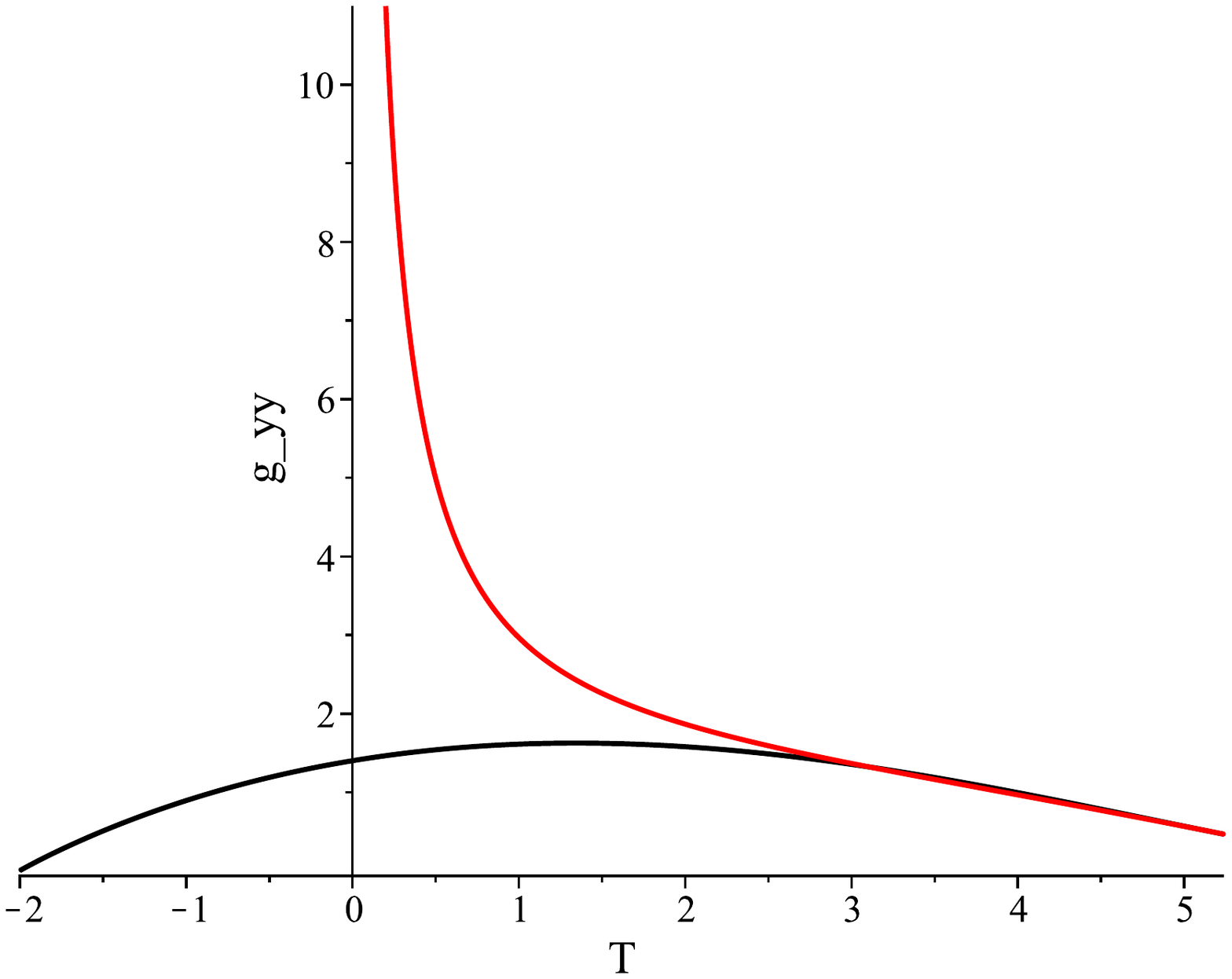}\\[-5.0cm]
\hspace{-0.3cm}\includegraphics[width=1.0\columnwidth, clip]{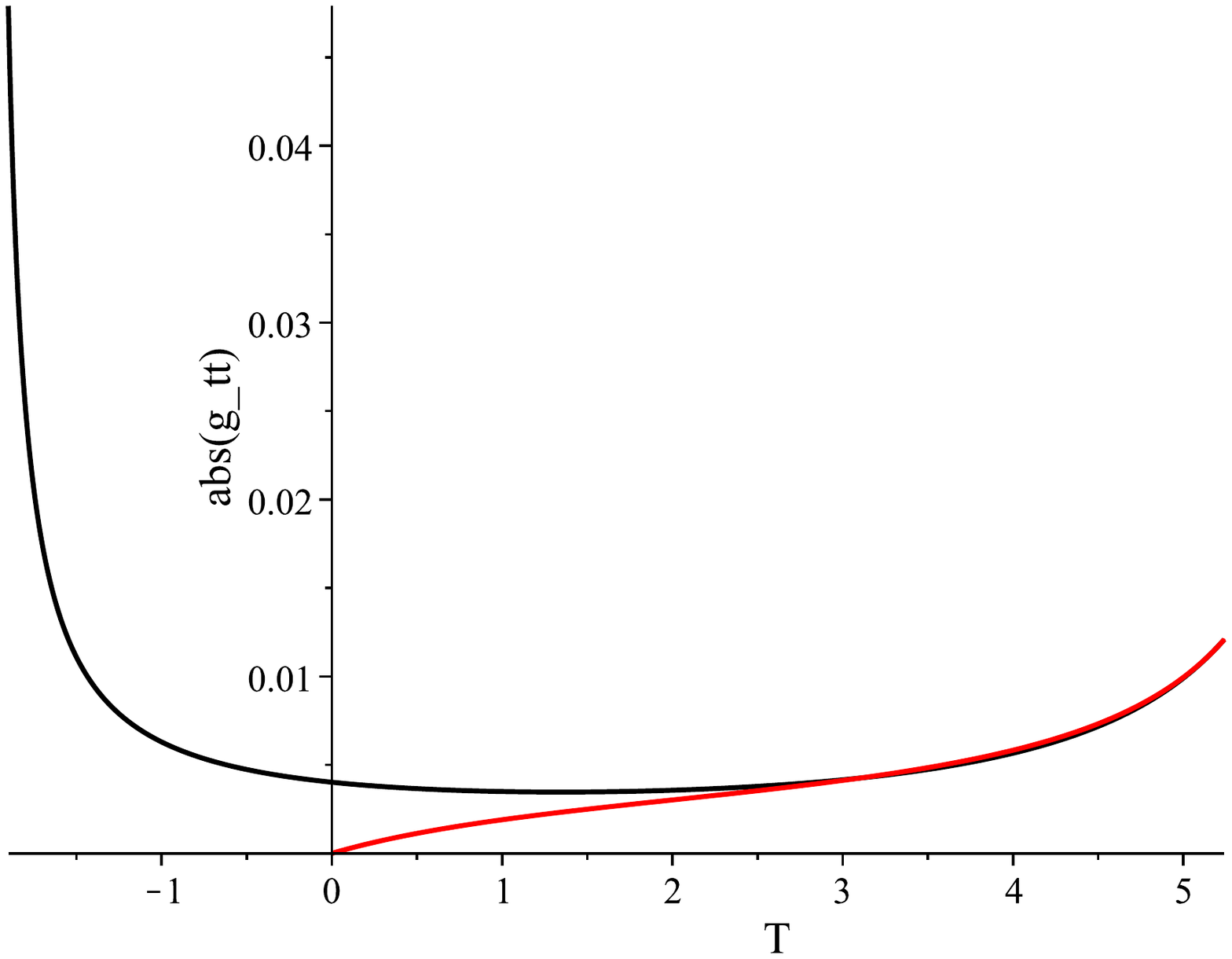}
\vspace{-5.0cm}\caption{{\linespread{0.4}{\footnotesize{$\theta\neq 0,\,\beta\neq 0$. com\-mut\-a\-tive (red) vs non-com\-mut\-a\-tive (black) interior of a higher-genus black hole. Top: The triad component $\Ethree$. Middle: The triad combination $(\Etwo)^2/\Ethree$, corresponding to $g_{yy}$. Bottom: $N^{2}$, corresponding to $|g_{\tau\tau}|$. Notice from the top graph that the radius of the 2D subspaces in this particular case does \emph{not} shrink to zero in the non-com\-mut\-a\-tive case, indicating the removal of the singularity, but the evolution stops due to $g_{yy}\rightarrow 0$ and $|g_{\tau\tau}| \rightarrow \infty$, indicating the presence of another horizon. The parameters are: $M=1$, $\Lambda=-0.1$, $\gamma=0.274$, $\theta=-0.05$, and $\beta=-0.1$.}}}}
\label{fig:thetabetahypds}
\end{figure}

The above analyses are somewhat complicated due to all the different possible scenarios. We therefore provide the following summary of all the possibilities in table 1:\\[0.0cm]

\clearpage
\begin{center}
{\vspace{-0.5cm}\large \underline{\small{Table 1}}}\enlargethispage{0.5cm}
\end{center}
{\small \begin{enumerate}\vspace{-0.5cm}
 \item {\bf Spherical topology:}
\begin{itemize}
\item $\theta\neq 0,\,\beta=0$: Small values of $\theta$ delay the singularity. Large values of $\theta$ remove the singularity.
\item ${\theta=0,\,\beta\neq 0}$: Singularity is removed.
\item ${\theta\neq 0,\, \beta\neq 0}$: Singularity is removed.
\end{itemize}

\item {\bf Toroidal topology}
\begin{itemize}
\item $\theta\neq 0,\,\beta=0$: Small values of $\theta$ delay the singularity. Large values of $\theta$ introduce a new  horizon$^{*}$.
\item ${\theta=0,\,\beta\neq 0}$: Singularity is removed.
\item ${\theta\neq 0,\, \beta\neq 0}$: Singularity is removed for large $\theta$. For small $\theta$ a new horizon appears$^{*}$.
\end{itemize}

\item {\bf Higher-genus topology}
\begin{itemize}
\item $\theta\neq 0,\,\beta=0$: Small values of $\theta$ delay the singularity. Large values of $\theta$ introduce a new  horizon$^{*}$.
\item ${\theta=0,\,\beta\neq 0}$: Singularity is removed.
\item ${\theta\neq 0,\, \beta\neq 0}$: Singularity is removed for large $\theta$. For small $\theta$ a new horizon appears$^{*}$.
\end{itemize}
\end{enumerate}
\vspace{-0.25cm}{\footnotesize $^{*}$ The presence of the new horizon prevents us from determining whether there is singular structure beyond the second  horizon.}}

\section{Concluding remarks}
In this manuscript we studied the effects of non-com\-mutative geometry on the interiors of black holes compatible with various topologies. The introduction of non-com\-mut\-a\-tiv\-ity was performed in two ways. In the first part of the study a smearing of the gravitating source was performed, mimicking the effects of the non-loc\-al\-iz\-a\-tion introduced by a non-trivial commutator between the spacetime coordinates. It was found that this smearing was capable of removing the curvature singularity in all scenarios. This result though is not that surprising, as one has essentially forced a smoothness onto the system. The resulting matter system, know in the literature as ``inspired by non-com\-mut\-a\-tive geometry'' will violate the energy conditions in the $T$-domain thus circumventing the results of the singularity theorems. However, it does hint at the fact that a possible resolution to the singularity issue lies in non-com\-mut\-a\-tive geometry effects.

In the second part of the study the Poisson algebra was directly altered by the introduction of a non-trivial bracket in i) the configuration degrees of freedom only, ii) the momentum degrees of freedom only, and iii) both. It was found that for some cases the singularity is merely delayed, occurring later (earlier in coordinate time) than in the corresponding com\-mut\-a\-tive scenario. However, in many cases some rather interesting results emerge. Either the singularity is removed, or else a new inner horizon forms. In the case of a new horizon, the domain that we are able to study with the method here is also non-singular. Overall, the presence of the parameter $\beta$ (non-trivial bracket between the triads) is more capable of singularity resolution than the parameter $\theta$ (non-trivial bracket between the connection). The results are summarized in table 1.

\vspace{0.7cm}
\section*{Acknowledgments}
AD would like to thank F. Rahaman and S. Ray (Kolkata) for useful discussions regarding noncom\-mut\-a\-tive geometry, and would like to acknowledge the Mathematics Department at Jadavpur University (Kolkata) for kind hospitality during these discussions. 

\PRLsep
\vspace{-0.080cm}

\linespread{0.7}
\bibliographystyle{unsrt}
 

%
%


} 
\end{document}